\documentclass[%
 reprint,
superscriptaddress,
preprintnumbers,
 amsmath,amssymb,
 aps,nofootinbib
]{revtex4-2}
\usepackage[english]{babel}
\usepackage{graphicx}
\usepackage{dcolumn}
\usepackage{bm}
\usepackage{hyperref}
\usepackage{color}
\usepackage[dvipsnames]{xcolor}
\usepackage{soul}
\usepackage{slashed}
\usepackage{tikz}
\usepackage{multirow}
\usepackage{cleveref}
\usepackage{comment}
\usepackage{ragged2e}
\usepackage{amsmath, amssymb}
\usepackage{siunitx}

\hypersetup{
    colorlinks,
    linkcolor={red!60!black},
    citecolor={blue!60!black},
    urlcolor={red!60!black}
}

 \usepackage{subcaption}
 \usepackage{qcircuit}
\usepackage{lettrine}

\input Zallman.fd

\newcommand{\ket}[1]{\left|#1\right\rangle}
\newcommand{\bra}[1]{\left\langle#1\right|}

  \makeatletter
    \renewcommand\@make@capt@title[2]{%
     \@ifx@empty\float@link{\@firstofone}{\expandafter\href\expandafter{\float@link}}%
      {\textbf{#1}}\@caption@fignum@sep#2\quad}%
    \makeatother
   
\makeatletter 
\renewcommand{\fnum@figure}{\textbf{Figure~\thefigure}}
\makeatother

\begin{document}


\title{Quantum Enhanced Dark-Matter Search with Entangled Fock States\\ in High-Quality Cavities}

\date{\today}
             
\author{Benjamin Freiman}%
\author{Xinyuan You}%
\author{Andy C. Y. Li}%
\author{Raphael Cervantes}%
\author{Taeyoon Kim}%
\author{Anna Grasselino}%
\author{Roni Harnik}%
\email{roni@fnal.gov}
\author{Yao Lu}%
\email{yaolu@fnal.gov}
\affiliation{Superconducting Quantum Materials and Systems (SQMS) Center, Fermi National Accelerator Laboratory, Batavia, IL 60510, USA}

\begin{abstract}
We present a quantum-enhanced protocol for detecting wave-like dark matter using an array of \(N\) entangled superconducting cavities initialized in an \(m\)-photon Fock state. By distributing and recollecting the quantum state with an entanglement-distribution operation, the scan rate scales as \(N^2(m{+}1)\) while thermal excitation is the dominant background, significantly outperforming classical single-cavity methods under matched conditions. We evaluate the robustness of our scheme against additional noise sources, including decoherence and beamsplitter infidelity, through theoretical analysis and numerical simulations. In practice, the key requirements, namely high-Q superconducting radio-frequency cavities that support long integration times, high-fidelity microwave beamsplitters, and universal cavity control, are already available on current experimental platforms, making the protocol experimentally feasible.
\end{abstract}

\maketitle


\section{Introduction}

Quantum devices and protocols are being harnessed to advance the search for dark matter~(DM). A well-motivated class of models that is amenable to quantum sensing is wave-like dark matter, where dark matter is assumed to be a light bosonic field that oscillates coherently at a frequency set by its mass. Such an interaction can resonantly displace fields in quantum devices that are frequency-matched to the DM. Examples of wave-like dark matter include the dark photon~\cite{Holdom:1985ag, Nelson:2011sf} and the axion~\cite{Peccei:1977hh, Preskill:1982cy}. In this class of dark matter exploration, the search proceeds as a scan over hypothetical dark matter masses and is limited by the time required to cover wide regions of DM mass. As a result, an important figure of merit for the search is the scan rate required to probe dark matter at a particular sensitivity. For example, a canonical axion and dark-photon search is ADMX~\cite{PhysRevLett.120.151301, ADMX:2019uok, PhysRevLett.134.111002}, which employs a tunable microwave cavity in a strong magnetic field and has set leading limits for $m_a/2\pi \approx 0.6-\SI{1.3}{\giga\hertz}$\footnote{We work in units with $\hbar=c=1$.}, a range that required multi-year data-taking to cover. Extending the search to higher frequency with this technology would suffer from slower progress due to the decreasing cavity size~\cite{Adams:2022pbo}. 

Quantum techniques can deliver transformative benefits in accelerating the scan rate of dark photon and wave-like dark matter searches. Recent theoretical investigations have identified entanglement operations as a valuable resource for coherently combining the feeble actions of the wave-like dark matter~\cite{Brady:2022qne, PhysRevLett.133.021801, Ito:2023zhp, Chen_2025, Bodas:2025vff}. Compared to an unentangled array, where the scan rate is typically linear with the number of sensors, $N$, the entangled array exhibits a favorable quadratic scaling, $\mathcal{R}_\textrm{scan}\propto N^2$. In addition, the detection of a DM weak signal traditionally relies on homodyne measurement and linear amplifiers, which are susceptible to quantum fluctuations, resulting in the standard quantum limit (SQL)~\cite{PhysRevD.88.035020}. Quantum-enhanced detection schemes, such as squeezing injection~\cite{Backes2021}, state-swapping and two-mode squeezing~\cite{PRXQuantum.2.040350, chen2022, PRXQuantum.4.020302, chen2025}, Schrödinger cat states~\cite{Zheng_2025},  microwave photon counting~\cite{PhysRevLett.126.141302, PhysRevLett.132.140801, PhysRevApplied.21.014043, braggio2025, PhysRevLett.131.211001, Kuo2025}, and other non-gaussian techniques~\cite{Gardner2025}, have been recently demonstrated to evade the SQL. In particular, qubit-assisted microwave photon counting mitigates quantum fluctuations by measuring only the photon number without vacuum fluctuation, making it fundamentally advantageous in achieving extremely low (sub-SQL) background rates~\cite{PhysRevD.88.035020}. The sensitivity is limited by the sub-percent-level residual thermal population in the detector. Photon counting can also benefit from stimulated emission~\cite{PhysRevLett.132.140801} that enhances the signal rate to $\mathcal{R}_s\propto m+1$ when the cavity is initiated in a non-classical Fock state $\left|m\right>$ and can correspondingly enhance the scan rate. To our knowledge, a protocol combining entanglement, photon counting, and stimulated emission has not yet been proposed. 
\begin{figure}[t]
 \begin{subfigure}[b]{0.49\textwidth}
    \includegraphics[width=\textwidth]{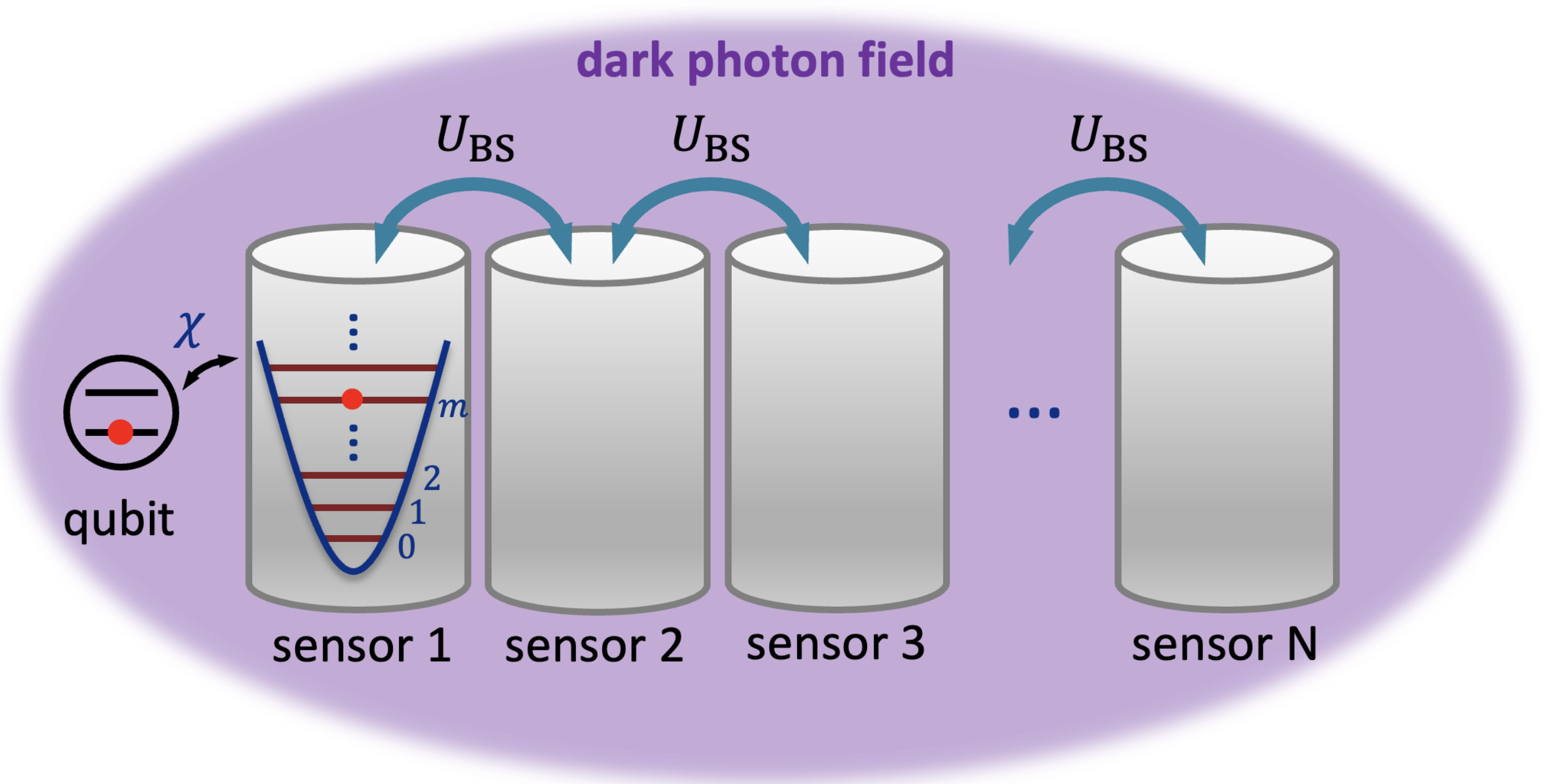}
        \label{fig:figure2}
\end{subfigure}
\captionsetup{justification=raggedright,singlelinecheck=false}
\caption{ \justifying A sketch of our proposed system. An array of $N$ cavities is networked by high-fidelity beamsplitter operations. The first cavity in the array is also coupled to a qubit for the intial Fock state preparation, and readout.}
\label{fig:schematic}
\end{figure}

\begin{figure*}[t]
    \begin{subfigure}[]{.65\textwidth}
        \[
        \begin{tikzpicture}
            \node at (0, 0) {
                \Qcircuit @C=3em @R=1em {
                    \lstick{\text{qubit}} & \ctrl{2}  & \qw & \qw & \qw & \meter & \multigate{5}{\text{RESET}} & \qw \\
                    \lstick{\text{}} & \text{}  & \text{} & \text{} & \text{} & \text{} & \text{} & \text{} \\
                    \lstick{\text{primary cavity}} & \gate{\lvert m \rangle_1} & \multigate{3}{U_\mathrm{ED}} & \gate{D_1(\alpha)} & \multigate{3}{U_\mathrm{ED}^\dagger} & \ctrl{-2} & \ghost{\text{RESET}} & \qw \\
                    \lstick{\text{cavity 2}} & \qw & \ghost{U_\mathrm{ED}} & \gate{D_2(\alpha)} & \ghost{U_\mathrm{ED}^\dagger} & \qw & \ghost{\text{RESET}} & \qw \\
                    \lstick{\vdots} & \vdots & & \vdots & & \vdots & \vdots \\
                    \lstick{\text{cavity N}} & \qw & \ghost{U_\mathrm{ED}} & \gate{D_N(\alpha)} & \ghost{U_\mathrm{ED}^\dagger} & \qw & \ghost{\text{RESET}} & \qw \\
                    & \text{\scriptsize State Prep.} & \text{\scriptsize Entang. Dist.} & \text{\scriptsize DM Int.} & \text{\scriptsize Inv. Entang. Dist.} & \text{\scriptsize Msmt.}  & \text{\scriptsize Reset} 
                    \gategroup{3}{3}{6}{5}{.7em}{--}
                }
            };
            \node at (-.90, 1.30) {\Large $\overbrace{\hspace{4.65cm}}^{\mathbf{D_1(\sqrt{N}\alpha)}}$};
        \end{tikzpicture}
        \]
        \label{fig:sub1}
    \end{subfigure}
\vspace{-1em}
\caption{\label{fig:protocol} \justifying
Quantum circuit for detecting DM with entangled Fock states. A prepared \(m\)-photon Fock state in the primary cavity is distributed across \(N\) cavities by the entanglement-distribution gate \(U_{\mathrm{ED}}\). During the integration time \(\tau_{\text{int}}\), the dark-photon field displaces each mode by \(D_i(\alpha)\). Applying \(U_{\mathrm{ED}}^{\dagger}\) coherently maps the uniform displacement to the primary cavity, \(D_1(\sqrt{N}\alpha)\), which is measured via the ancilla qubit, followed by reset.}
\end{figure*}
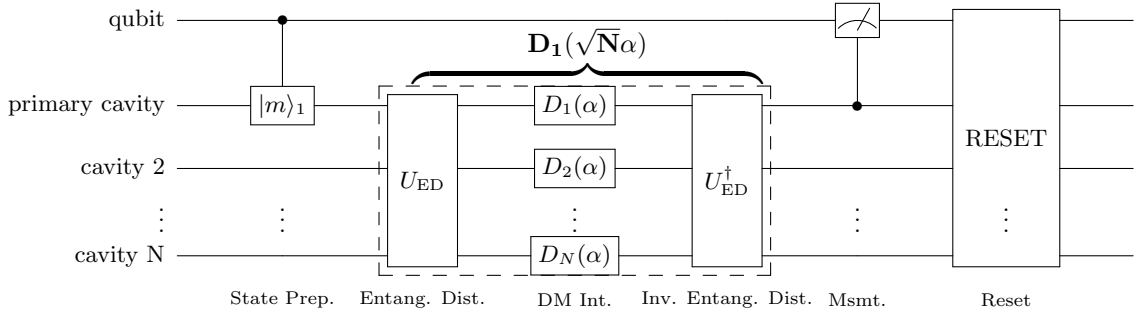

In this Letter, we introduce a quantum-enhanced protocol to accelerate the search for wave-like dark matter, using dark photons as an example. Our method employs a network of high-quality-factor superconducting cavities, interconnected via microwave beamsplitter operations, as depicted in Fig.~\ref{fig:schematic}. By utilizing entangled Fock states distributed across the sensor array, the protocol harnesses the coherence of the dark matter field and amplifies the signal via stimulated emission. We demonstrate a significant quantum enhancement in the search scan rate, scaling favorably with the number of cavities and the initial Fock state photon number, in the presence of realistic noise sources such as thermal heating, sensor mode decoherence, and beamsplitter infidelity. Through theoretical analysis and simulations, we establish the combined advantages of entanglement and stimulated emission, offering a practical and scalable approach for future dark matter experiments.

\section{the protocol}
Our quantum-enhanced protocol, using an array of superconducting cavities controlled by an ancillary qubit, is depicted in Fig.~\ref{fig:protocol}. It harnesses entanglement and stimulated emission to improve the scan rate over classical single-cavity approaches, which involves five key steps: (1) A single cavity is prepared in an m-photon Fock state, with others in the vacuum state; (2) a series of unitary beam-splitter pulses, dubbed an entanglement distribution (ED) gate, symmetrically distributes the state among all cavities; (3) the array interacts with the dark photon field over a set time $\tau_{\text{int}}$, accumulating coherent displacements in the cavities; (4) an inverse ED gate is applied to concentrate the signal into one cavity for amplification; and (5) measurement and reset are performed to detect the signal and prepare for the next cycle.

Our protocol benefits directly from long-coherence sensors and high-fidelity entanglement operations. High-Q SRF cavities~\cite{Kim2025} reduce decoherence-induced backgrounds, extending the useful integration window and increasing both SNR and scan rate. Likewise, tunable microwave beamsplitters with high fidelities~\cite{lu2023high, chapman2023} preserve the ED mapping with minimal overhead. Together, these ingredients enable us to fully exploit the protocol’s quantum advantage.

As we shall see in later sections, this enables a quantum enhancement of the scan rate in the search for dark photon dark matter. This enhancement is underpinned by the protocol’s key features: the signal rate scales quadratically with the number of cavities in the network, and is further amplified by the initial Fock state number through stimulated emission. Notably, the dominant incoherent thermal noise accumulated during integration remains independent of the cavity number, matching the noise level of a single cavity. Moreover, this noise can be characterized in situ by photon counting across multiple cavities in the network, up to corrections from dephasing.

\section{Enhanced signal population}
\subsection{Coherent interactions of Wavelike Dark Matter}
The dark photon is a hypothetical new vector particle that mixes with the standard model photon through its kinetic interactions and has a mass~$m_{A'}$~\cite{Holdom:1985ag}. The dark photon can also make up dark matter and is an attractive target for direct searches, which employ its mixing with the standard model photon~\cite{Antypas:2022asj}. The kinetic mixing interaction can be written as a modification to the Hamiltonian $H \supset \epsilon \vec{E}\cdot\vec{E'}$, where $\vec{E}$ and $\vec{E'}$ are the electric fields associated with the Standard Model and dark photons, respectively, and $\epsilon$ is the kinetic mixing parameter. The dark photon field, if it constitutes cold dark matter, oscillates as a classical coherent state with a frequency set by the dark photon mass and an amplitude determined by its energy density~$\rho_\mathrm{DM}$. The dark matter field can then act as a source of photons in electromagnetic devices, such as cavities.
The physical size of the cavity is set by the inverse frequency of the dark photon, $L_{\text{cav}} \sim \omega_\mathrm{DM}^{-1}$. Since the dark photon de Broglie wavelength is much greater than the wavelength $L_{\text{DM}} \sim 10^3 \omega_{\text{DM}}^{-1}$ is much greater than the cavity size, a sensor network where cavities are placed close to each other will be subject to a coherent dark photon signal that displaces all cavities coherently~\cite{Brady:2022qne}.

In this work, we consider an array of $N$ cavities tuned to the same frequency, each with a volume of $V_\mathrm{cav}\sim L_\mathrm{cav}^3$. As shown in Appendix~\ref{app:dark photon interaction}, so long as the system is fully within a DM coherence length, it will displace all cavities in phase, and the DM-cavity mode interaction, in the interaction picture and under the rotating-wave approximation, is
\begin{equation}
    H_\textrm{int}= g  
     e^{i\Delta  t+i\delta\phi(t)}
     \sum_{n=1}^N
     a_n^\dagger +\mathrm{h.c.}.
     \label{H_DP}
\end{equation}
Here, $a^\dagger_n$ is the creation operator in the $n$-th cavity, and $\Delta = \omega-m_{A'}$ is the detuning between the cavity frequency and dark photon frequency. 
The coupling of the dark matter field to the cavity mode is $g = \epsilon\,  \sqrt{2 G\rho_\mathrm{DM} V_\mathrm{cav}\, \omega}$, and $G$ is the cavity form factor of order one, which is defined in Appendix~\ref{app:dark photon interaction}.
The phase $\delta\phi(t)$ is the random, slowly varying phase of the dark-matter field. It evolves on a coherence time $\tau_{\mathrm{DM}}\simeq 10^{6} m_{A'}^{-1}$ set by the field’s velocity dispersion, which corresponds to an effective coherence quality factor $Q_{\mathrm{DM}}\sim 10^{6}$. As a result of the interaction in Eq.~\eqref{H_DP}, the cavity field undergoes
a random walk process over the integration time $\tau_\textrm{int}$, leading to a stochastic displacement $D_i\left(\alpha(\tau_\mathrm{int})\right)$ coherent across all of the cavities in the network, with the phase correlation time of the displacement $\alpha$ set by $\tau_\mathrm{DM}$. In the resonance case, the ensemble-averaged displacement amplitude is given by (see Appendix~\ref{app:random_walk})
\begin{equation}
\langle\lvert\alpha(\tau_\mathrm{int})\rvert\rangle = 
    \sqrt{2 |g|^2 \tau_\mathrm{DM} \left[ \tau_\mathrm{int} - \tau_\mathrm{DM} \left(1 - e^{-\frac{\tau_{\text{int}}}{ \tau_\mathrm{DM}}}\right) \right]}\,.\label{eq:DM_displacement}
\end{equation}
Here, $\langle\boldsymbol{\cdot}\rangle$ represents ensemble averaging over DM's phase fluctuation. In the limit of large detuning, the displacement is suppressed by a Lorentzian factor. We assume the sensing device to be within the near-resonance regime, $\Delta\lesssim \tau_\textrm{DM}^{-1}$, where the displacement is well represented by Eq.~\eqref{eq:DM_displacement}.

In the next two sections, we show how the DM field’s spatial coherence produces an effective DM-cavity interaction rate that scales with the number of cavities, which can be further combined with a higher initial photon number to increase the rate.

\subsection{Cavity-number-enhanced Displacement}

\begin{figure*}[t!]
\centering
\includegraphics[width=0.95\textwidth]{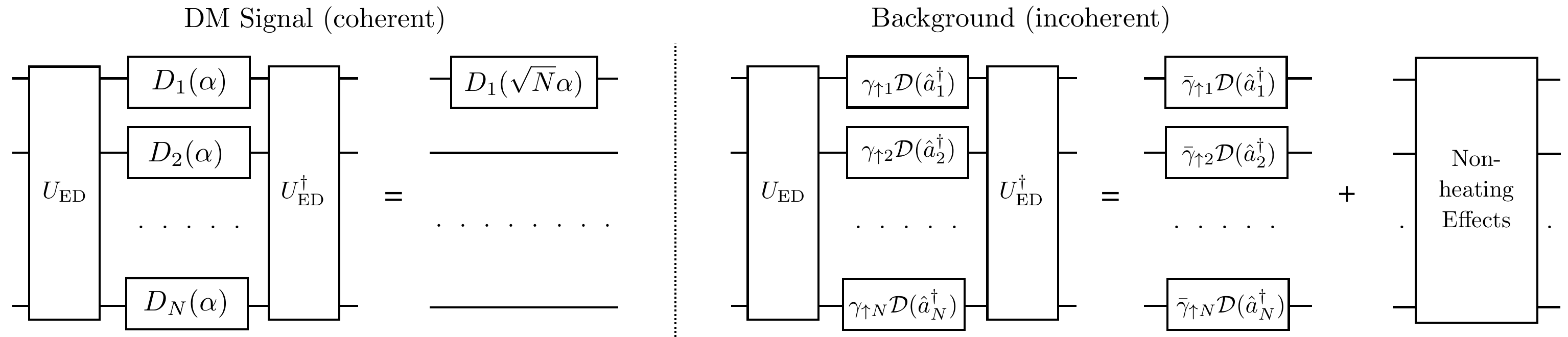}
\caption{\justifying
Effective circuit relations for the dark matter signal~(left) and cavity heating during integration time~(right), which is the dominant source of noise. The signal is coherent across the cavity network and thus scales with the number of cavities $N$, whereas the background is incoherent and does not scale. The non-heating effects can arise from photon loss \(\gamma_{\downarrow}\), dephasing \(\gamma_{\phi}\), and beamsplitter infidelity, which reduce sensing efficiency but do not introduce \(N\)-scaling of the background.}

\label{fig:circuits-eqns}
\end{figure*}
Because dark matter displaces the collective mode in a superposition of all cavities (see Eq.~\eqref{H_DP}), we are motivated to consider a unitary operation that transfers a photon from a primary signal cavity to such a superposition,
\begin{equation} 
\label{eq:ForwardEtnanglement}
U_\mathrm{ED}\hat{a}^\dagger_1 
   U_\mathrm{ED}^{-1} 
   = \frac{1}{\sqrt{N}}\sum_{n=1}^N  \hat{a}_n^\dagger.
\end{equation}
We call this transformation the entanglement distribution (ED). As defined above, ED gates are not unique, and  
can be achieved by a sequence of beamsplitter transformations (see Appendix~\ref{app:Entanglement-enhanced-displacement}).
The inverse ED takes an equal superposition and transfers it to the primary cavity
\begin{equation} \label{eq:InverseEtnanglement}
   U_\mathrm{ED}^{-1} \sum_{n=1}^N \hat{a}^\dagger_n U_\mathrm{ED} = \sqrt{N} \hat{a}^\dagger_1.
\end{equation}
Using this relation, it is simple to prove the following: 
a coherent displacement of all $N$ cavities sandwiched between an ED gate and an inverse-ED produces an enhanced displacement of the signal cavity, 
\begin{equation} 
U_\textrm{ED}^{-1}D_1 (\alpha) \otimes  D_2 (\alpha)  \otimes ... \otimes D_N  (\alpha)U_\textrm{ED} \\
   = D_1(\sqrt{N}\alpha)\,,
\label{eq:disp_enhancement}
\end{equation}
The proof is shown in Appendix~\ref{app:Entanglement-enhanced-displacement}. This equivalence is depicted as the quantum circuits in the left panel of Fig.~\ref{fig:circuits-eqns}. In the weak-displacement limit, a subsequent photon-counting test gains an
$N$-fold increase in transition probability relative to a single cavity.

\subsection{Stimulating the Signal with Fock States} 

We have shown that our sequence of (ED)-(DM-induced displacement)-(ED)$^{-1}$ yields an equivalent $\sqrt{N}$-enhanced displacement of the primary cavity. Since this holds at the unitary level, it is independent of the cavity's initial state. In particular, we can choose to further enhance the signal rate by starting with the $m$-th Fock state as in ~\cite{PhysRevLett.132.140801}. Because creation operators satisfy $a^\dagger |m\rangle = \sqrt{m+1}|m+1\rangle$, the $m\!\to\! m+1$ transition amplitude scales as $\sqrt{m+1}$, so the transition rate is enhanced by a factor of $m+1$. As a result, after an integration time $\tau_{\mathrm{int}}$, the probability that DM induces a transition in the first cavity from $|m\rangle$ to $|m{+}1\rangle$ is 
\begin{equation}
\begin{aligned}
    &n_{s}^0(\tau_\textrm{int}) = \langle|\bra{m+1}_1\mathcal{D}_1 (\sqrt{N}\alpha(\tau_\textrm{int}))\ket{m}_1|^2\rangle \\
    & \approx \langle|\bra{m+1}_1\sqrt{N}\alpha(\tau_\textrm{int}) \hat{a}_1^\dagger \ket{m}_1|^2\rangle = \langle\lvert\alpha(\tau_\textrm{int})\rvert\rangle^2 N(m+1)\,, \label{eq:signal_population}
\end{aligned}
\end{equation}
which combines the enhancement from both the number of cavities~$N$ and the initial Fock number~$m$.
While the DM-induced transition to $|m-1\rangle$ also exists (unless we start with $m=0$), we adopt the presence of $m+1$ photons as the proxy of DM signal~\cite{PhysRevLett.126.141302}, so that the false positive rate is not dominated by the cavity decay, but by the much weaker cavity heating instead. 

\section{Signal, Noise, and SNR 
} 

In this section, we discuss the rates for signal and for false positives, and combine them to get a signal-to-noise ratio (SNR). We will first discuss the relevant quantities in the limit of no loss (except for weak cavity heating), and then consider other loss effects, such as cavity decay and dephasing. \\[-5pt]

\paragraph*{Signal Rate:}
Combining Eq.~\eqref{eq:DM_displacement} and Eq.~\eqref{eq:signal_population}, we obtain the DM-induced displacement of the cavity field as a time-dependent probability rate over integration, as
\begin{align}
\label{eq:Rs}
\mathcal{R}^0_s(\tau_\text{int}) &= \frac{ n_{s}^0(\tau_\text{int})}{\tau_{\text{int}}}  \\ \nonumber
&= 2 |g|^2 \tau_\mathrm{DM} \left[ 1 - \frac{\tau_\mathrm{DM}}{\tau_{\text{int}}} \left(1 - e^{-\frac{\tau_{\text{int}}}{ \tau_\mathrm{DM}}}\right) \right] N(m+1).\label{eq:Rs0}
\end{align}
In the long and short limits of the integration time relative to $\tau_\mathrm{DM}$, we obtain
\begin{equation} \label{eq:ballistic_to_diffusive}
\mathcal{R}^0_s(\tau_\text{int}) \approx 
\begin{cases} 
|g|^2 \tau_{\text{int}} N(m+1) & \text{if } \tau_{\text{int}} \ll 
 \tau_{\text{DM}}, \\
2 |g|^2 \tau_\mathrm{DM} N(m+1) & \text{if } \tau_{\text{int}} \gg \tau_{\text{DM}}.
\end{cases}
\end{equation}
We see a coherently growing rate at short times, and a constant per-shot rate, for long integration times, as expected.
Here, we assume lossless cavities. In the presence of loss, the signal rate eventually drops due to depletion of the population at long integration times, which necessitates an optimization of the integration time. We will account for this effect at the end of this section.\\[-5pt]
\paragraph*{False Positives from Heating during Integration}:
Having established the enhancement of the signal in our system, we now turn to the noise in the search. A dominant source of false positives is thermal heating of the cavities during the integration. Such events occur independently and incoherently in the different cavities. At the circuit level, heating in cavity \(n\) is modeled by the Lindblad term \(\gamma_{\uparrow,n}\,\mathcal{D}[\hat a_n^{\dagger}]\), with \(\gamma_{\uparrow,n}\) the local heating rate and \(\mathcal{D}\) the standard dissipator.

Heating that occurs during the integration time, which dominates the cycle time, transforms into an effective heating rate across the network under the ED and inverse ED operations. This mapping is shown in the right panel of Fig.~\ref{fig:circuits-eqns}. Since the signal is defined as the appearance of the $(m+1)$th photon in cavity~1, the corresponding false-positive (background) rate is
\begin{equation}
\mathcal{R}_b^0\sim (m+1) \bar \gamma_{\uparrow 1}\,,
\label{eq:BG-rate}
\end{equation}
where the factor $m+1$ accounts for bosonic stimulation of the $m\to m+1$ transition.
The effective rates $\bar \gamma_{\uparrow,n}$ are related to the physical heating rates $\gamma_{\uparrow,n}$ via the inverse-ED transformation, which is designed to collect the equal-weight (symmetric) mode into the primary cavity. Hence, the effective rate for false positives in cavity~1 is (see Appendix~\ref{app:parasitic1} for detailed derivation)
\begin{equation}
\bar\gamma_{\uparrow 1} = \frac{1}{N}\sum_{n=1}^N \gamma_{\uparrow,n}\,,
\label{eq:gamma-bar1}
\end{equation}
i.e., the average single-cavity heating rate. We see that \emph{while the signal rate is enhanced by $N$, the dominant false-positive rate remains that of a single cavity}.


\paragraph*{Signal-to-Noise Ratio:}
Repeating the protocol many times to accumulate a total integration time $\tau_\mathrm{tot}$,
Eqs.~\eqref{eq:BG-rate} and \eqref{eq:gamma-bar1} imply an \emph{expected} number of false positives
$R_b\,\tau_\mathrm{tot}$. If this background rate is known precisely (from calibrated modeling or from
independent off-resonance measurements), the sensitivity is limited by the statistical fluctuation
of the background counts. For Poisson-distributed counts one has
$\mathrm{Var}[n_b^0]=R_b^0\tau_\mathrm{tot}$, and for large counts the distribution is well approximated
by a Gaussian with standard deviation $\delta(R_b^0\tau_\mathrm{tot})=\sqrt{R_b^0\tau_\mathrm{tot}}$.
The resulting signal-to-noise ratio is
\begin{equation}
\label{eq:SNR}
    \mathrm{SNR}^0=\frac{R_s^0\,\tau_\mathrm{tot}}{\sqrt{R_b^0\,\tau_\mathrm{tot}}}
    \sim |g|^2\,\tau_\mathrm{DM}\,N\,\sqrt{(m+1)\,\tau_\mathrm{tot}\,\bar\gamma_{\uparrow 1}^{-1}},
\end{equation}
assuming the long integration time limit $\tau_\textrm{int}\gg \tau_\textrm{DM}$.

Before proceeding, we note that within our protocol, one can \emph{measure} the false positive rate in situ, by performing measurements on other cavities in the network. In Appendix~\ref{app:simultaneous_noise_measurement}, we show that $\bar\gamma_{\uparrow 1} = \frac{1}{N-1}\sum_{i=2}^N \bar \gamma_{\uparrow i}$, namely, that the false positive rate in the signal cavity equals the average noise rate in all other cavities. This is a consequence of the unitarity of $U_\mathrm{ED}$. Consequently, the false positive rate may be measured independently of, and simultaneously with, the search itself. A similar strategy with a different interpretation was proposed in~\cite{shu2024eliminatingincoherentnoisecoherent}.

\begin{figure}[t!]
\centering
\includegraphics[width=0.4\textwidth]{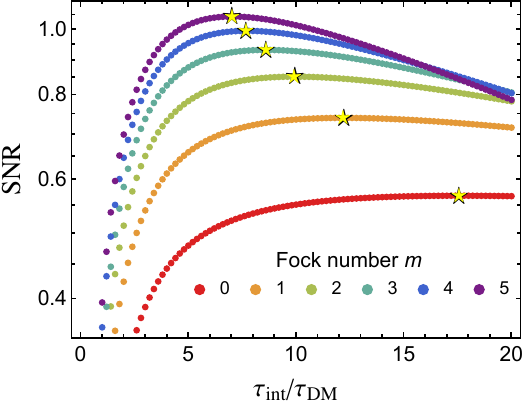}
\caption{\justifying 
The dependence of the signal-to-noise ratio on the single-shot integration time $\tau_\mathrm{int}$, for different choices of the initial Fock state $|m\rangle$, in a two-cavity setup. The cavities are at \SI{7}{\giga\hertz}, with decay time $T_\downarrow = \SI{2.27}{\milli\second}$ (about 100 times $\tau_\textrm{DM}$), dephasing time $T_\phi = \SI{22.7}{\milli\second}$, and a heating time of $T_\uparrow = \SI{1.9}{\second}$ (corresponding to an equilibrium temperature of \SI{50}{\milli\kelvin}). The SNR is evaluated for a total integration time of $\tau_\textrm{tot}= \SI{1}{\second}$, assuming an DM-cavity coupling of $g= \SI{73.6}{\hertz}$. Stars mark the optimal $\tau_{\text{int}}$ that maximizes the SNR, which shifts to shorter values as $m$ increases.}
\label{fig:SNR_vs_m_main}
\end{figure}
\paragraph*{Including Loss and other parasitic effects}
We now include the effects of decay in the cavity, as well as dephasing and beamsplitter infidelity. In the presence of decay, the initial Fock state $|m\rangle$ will drop to $|m-1\rangle$ at long integration times, drastically reducing the probability for the signal to excite the $|m+1\rangle$ state. Dephasing noise transformed by the entanglement distribution operations also leads to incoherent swapping of the primary cavity population to other cavities (see Appendix~\ref{app:parasitic1}). The integration time thus needs to be optimized, accounting for the decoherence time scales of the system and the DM, as well as the initial state $m$. In Fig.~\ref{fig:SNR_vs_m_main} we show the SNR as a function of the integration time (in units of $\tau_\mathrm{DM}$) for various values of $m$.
We observe two important features in this plot: (1) the SNR does not monotonically increase over the integration time, but reaches a maximum SNR before it falls down. (2) The optimal integration time corresponding to the maximum SNR is dependent on the Fock number $m$, and becomes shorter as $m$ increases. These features are natural consequences of the competition between coherent DM-induced build-up, which saturates once $\tau_{\text{int}}$ exceeds $\tau_{\rm DM}$, and $(m{+}1)$-enhanced loss and backgrounds that continue to accumulate (see Appendix~\ref{app:optimal_integration_time}).

Another important parasitic effect is beamsplitter infidelity in the entanglement–distribution operation. We model this by adding, only during the ED and inverse-ED windows, elevated cavity dephasing and decay on the coupled modes that reproduce a target single-photon swap fidelity $F_{\mathrm{BS}}$, and we propagate the full cycle with a Lindblad master equation simulation; see Appendix~\ref{app:simulation} for details. We find that the SNR decreases with beamsplitter infidelity, and larger $m$ makes the distributed state more sensitive. However, increasing $\tau_{\mathrm{int}}$ helps dilute this penalty by making the fixed ED overhead a smaller fraction of each cycle, a benefit enabled by high-Q cavities that allow a longer optimal $\tau_{\mathrm{int}}$. For beamsplitter fidelities at or above $99\%$, which are well-achievable in circuit-QED platforms~\cite{lu2023high,chapman2023, li2025, Kim2025}, the maximum SNR remains close to the ideal scaling, confirming that the protocol is robust to realistic beamsplitter imperfections.

\section{Scan rate and Sensitivity}
\begin{figure}[t]
\includegraphics[width=0.42\textwidth]{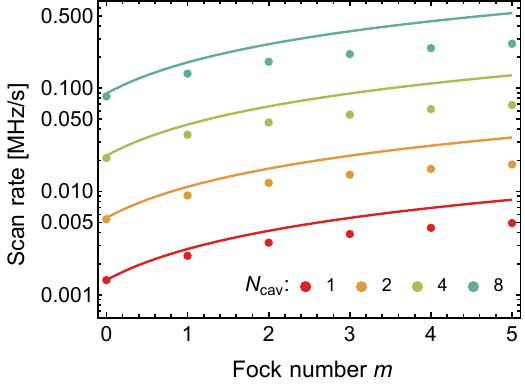}
\captionsetup{justification=raggedright,singlelinecheck=false}
\caption{\justifying
Scan rate at 90\%~CL exclusion versus the number of cavities $N$ for several initial Fock numbers $m$, normalized to the single-cavity rate. Solid lines show the ideal prediction from Eq.~\eqref{eq:scan_rate} without loss. Dots represent Lindblad master equation simulations at optimally chosen $ \tau_\text{int} $, with beamsplitter fidelity $\mathcal{F}_\textrm{BS}=99$\% and other parameters the same as Fig.~\ref{fig:SNR_vs_m_main}.
}
\label{fig:scan_rate_vs_mn}
\end{figure}

\begin{figure*}[t!]
\includegraphics[width=0.8\textwidth]{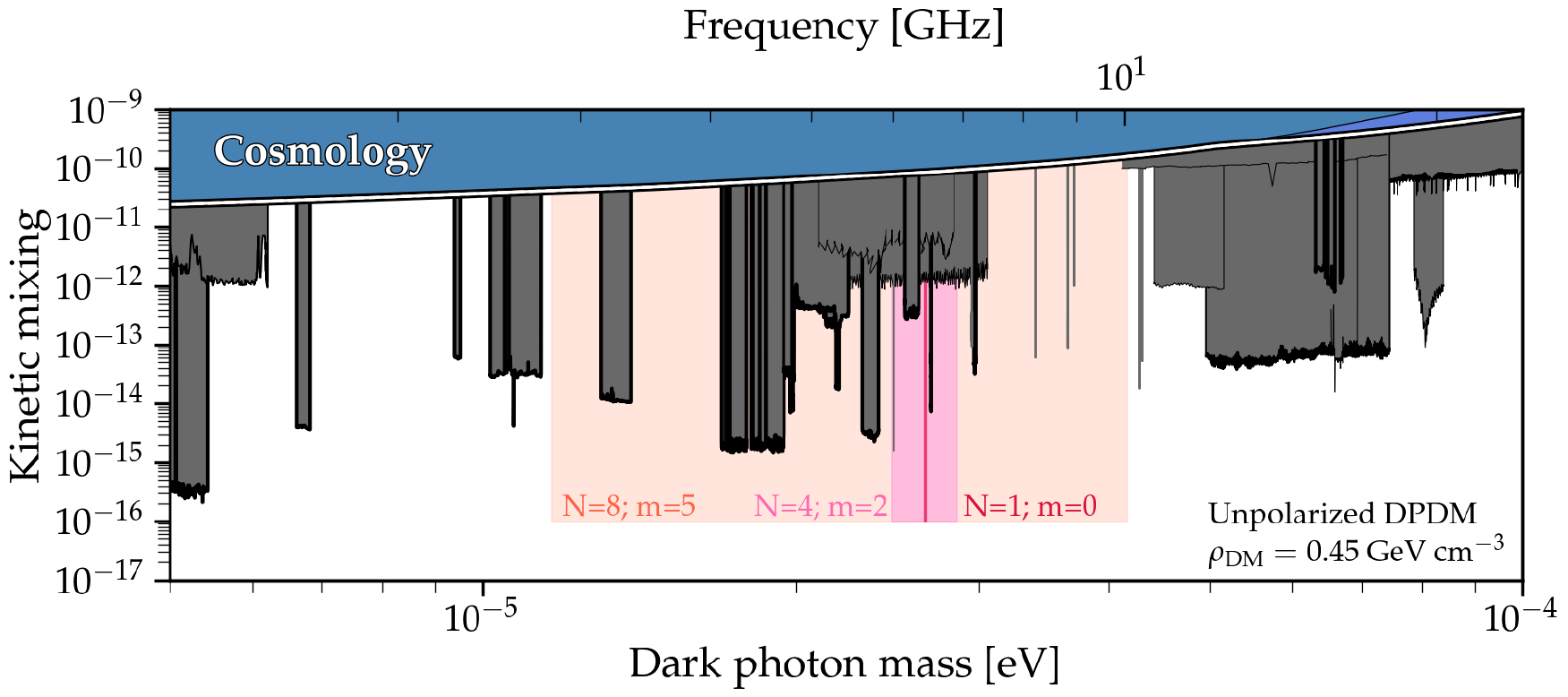}
\label{fig:figure2}
\caption{\justifying Expected 90\% CL reach for dark-photon dark matter. Shaded pink bands show the frequency range that can be covered in a 15-hour scan at target coupling $\epsilon = 10^{-16}$ for three $(N,m)$ configurations. We assume unpolarized DPDM with $\rho_{\rm DM}=0.45~\mathrm{GeV\,cm^{-3}}$. The other limits from direct searches and cosmology compiled in~\cite{Caputo:2021eaa} are also shown.}
\label{fig:reach}
\end{figure*}
Because resonant searches for wave-like DM scan across frequency, the natural figure of merit is the scan rate $\mathcal{R}_\mathrm{scan}$, defined as the DM mass bandwidth that can be probed per measurement, down to the target coupling $\epsilon$. Each step in the scan covers a bandwidth of order $\tau_\mathrm{DM}^{-1}$ in DM mass parameter space~\cite{Cervantes2025}. The total time per step $\tau_\mathrm{tot}$ at frequency $\omega$ follows from imposing a target SNR $\zeta_\textrm{SNR}$ in Eq.~\eqref{eq:SNR} for the target $\epsilon$.
Then, the resulting scan rate can be expressed as (see Appendix~\ref{app:projected_sensitivity})
\begin{align}
    \mathcal{R}_\mathrm{scan} &=\tau_\textrm{DM}^{-1}\tau_\textrm{tot}^{-1}\nonumber\\
    &\sim
    \eta^2\epsilon^4 G^2\rho^2_\mathrm{DM} V^2_\mathrm{cav}\, \omega^2\zeta^{-2}_\text{SNR}n^{-1}_\text{th}\tau_\text{cav}\tau_\text{DM}N^2(m+1)\label{eq:scan_rate}
\end{align}
which is a factor of $N (m+1)$ faster than a sensor composed of $N$ independent cavities in the vacuum state. Here, $\rho_{\mathrm{DM}} = \SI{0.45}{\giga\electronvolt\per\cubic\centi\metre}$ is the density of the dark matter, $n_\textrm{th}$ is the cavity thermal occupation, and $\eta$ is the efficiency factor that captures all parasitic effects during a cycle, including cavity decay, dephasing, and the beamsplitter infidelity. In the ideal limit where $\eta=1$, Eq.~\eqref{eq:scan_rate} gives the solid curves in Fig.~\ref{fig:scan_rate_vs_mn} for different Fock state number $m$ and cavity number $N$, assuming cylindrical cavities of $G=0.23$ and $V_\textrm{cav} = $ \SI{55.5}{\cubic\centi\metre}. When the parasitic effects are included, the scan rates are given by the dots in the same figure, obtained from Lindblad master equation~\cite{Lindblad1976} simulations at the corresponding optimal $\tau_\textrm{int}$. Despite efficiency losses, the entangled Fock-state approach retains a clear advantage under realistic noise, yielding an enhancement of $\sim 300$ when moving from a single-cavity vacuum sensor to an eight-cavity entangled array prepared in a five-photon Fock state.

The potential reach with an entangled cavity array is shown in Fig.~\ref{fig:reach} as 90\%~CL exclusion, assuming tunable cavities. We plot three $(N,m)$ configurations as shaded bands, using the same noise parameters as in the previous simulations. For a fixed target coupling $\epsilon = 10^{-16}$ and a fixed 15-hour run, we compute at each probe frequency $\omega$ the required per-step exposure $\tau_{\mathrm{tot}}(\omega)$ from the SNR condition, and, with a step size set by $\tau_{\mathrm{DM}}^{-1}$, accumulate the covered bandwidth until the time budget is exhausted. The width of each band is the resulting scan range. This comparison clearly shows the advantage of larger $N$ and $m$, which translates into much faster scanning of DM.

\section{Discussion and Outlook}

In this work, we propose a novel DM detection scheme based on high-Q SRF cavities, beamsplitters, and photon counting. The sensing protocol provides a signal rate that scales with $N$, the number of cavities in the network, and we have shown that the dominant thermal background does not scale with $N$. The protocol further enables an enhancement of the signal rate by stimulated emission, where a Fock state of $m$ photons provides an additional boosting factor of $m+1$. The resulting scan rate scales as $\mathcal{R}_\textrm{scan} \propto N^2(m+1)$ in the ideal case. We have studied the protocol through a comprehensive simulation that accounts for realistic noise sources, such as cavity decay, dephasing, and beamsplitter infidelity, to validate our results. The required ingredients, namely long cavity coherence and high-fidelity control operations, have been demonstrated on current experimental platforms \cite{lu2023high, Kim2025}. A full search further calls for an efficient tuning mechanism for the cavity network, which can be implemented with mechanical elements \cite{Read2023} or with parametrically tunable superconducting circuits \cite{Zhao:2025thg, Zheng_2025}.

\section{Acknowledgments}
We would like to thank Zachary Goff-Eldredge, Srivatsan Chakram, Daniel Molenaar, Arushi Bodas, Sohitri Ghosh, Bianca Giaccone, Alex Millar, Ciaran O'Hare, Jens Koch, Gianpaolo Carosi, and David I. Schuster for their valuable discussions. This work was supported by the U.S. Department of Energy, Office of Science, National Quantum Information Science Research Centers, Superconducting Quantum Materials and Systems Center (SQMS), under Contract No. 89243024CSC000002. Fermilab is operated by Fermi Forward Discovery Group, LLC under Contract No. 89243024CSC000002 with the U.S. Department of Energy, Office of Science, Office of High Energy Physics.
YL acknowledges support from the DOE Early Career Research Program. RH was also supported by the DOE, Office of Science, Office of High Energy Physics, under the Quantum Information Science Enabled Discovery (QuantISED) program. 

\textit{Data availability}---The data presented in this study are openly available at \href{https://doi.org/10.5281/zenodo.18296589}{10.5281/zenodo.18296589}.

\bibliographystyle{apsrev4-2}
\bibliography{bib}

\onecolumngrid
\newpage
\appendix
\setcounter{equation}{0}
\setcounter{figure}{0}

\makeatletter
\@removefromreset{equation}{section}
\@removefromreset{figure}{section}
\makeatother

\renewcommand{\theequation}{S\arabic{equation}}
\renewcommand{\thefigure}{S\arabic{figure}}

\renewcommand{\theHequation}{S\arabic{equation}}
\renewcommand{\theHfigure}{S\arabic{figure}}

\section{Dark Photon Dark Matter and its Interaction with Cavity Modes}
\label{app:dark photon interaction}

In this appendix section, we review dark photon dark matter and its interaction with an electromagnetic cavity. 
The kinetic mixing interaction can be written as a perturbation to the cavity Hamiltonian $H_0$,
\begin{equation}
    H = H_0 + \epsilon \vec{E}\cdot\vec{E'}+\ldots,\label{eq:EdotE'}
\end{equation}
where $E$ and $E'$ are the electric fields associated with the Standard Model and dark photons, respectively.  The Standard Model electromagnetic field can be expanded in cavity modes. Focusing only on a specific cavity mode at frequency $\omega$ with field profile $\hat E$, the field can be expressed as 
\begin{equation}
    \vec E = \sqrt{\frac{\omega}{2V_\mathrm{cav}}} \hat E_{\vec x} a^\dagger e^{i\omega t} +\mathrm{h.c.}\label{eq:E}
\end{equation}
The dark photon field, if it constitutes cold dark matter, oscillates as a classical coherent state with a frequency set by the dark photon mass $m_{A'}$ and an amplitude determined by its density 
\begin{equation}
   \vec E' =
  \sqrt{2 \rho_\mathrm{DM}}\cos\left( m_{A'}t + \phi(t)\right)\,. \label{eq:E'}
\end{equation}
The phase $\phi(t)$ changes slowly, on a time scale set by the squared velocity dispersion of dark matter, of order $\sim 10^6 m_{A'}^{-1}$. The phase also changes in space on a scale of the dark matter de Broglie wavelength $\sim 10^3 m_{A'}^{-1}$. We note that the size of a cavity is of order $L_\mathrm{cav}\sim \omega^{-1} \sim m_{A'}^{-1}$. As a result, the dark matter is spatially coherent over a local network of cavities. 

Combining Eqs.~\eqref{eq:EdotE'}--\eqref{eq:E'}, the DM-mode interaction Hamiltonian is
\begin{eqnarray}
    H = 
     \omega \left(a^\dagger a +\frac{1}{2}\right) + g \cos\left( m_{A'}t + \phi(t)\right) 
     a^\dagger +\mathrm{h.c.},
\end{eqnarray}
where $a^\dagger$ and $a$ are creation and annihilation operators of the cavity mode. 
The coupling of the dark matter field to the cavity mode is
\begin{equation}
    g = \epsilon\,  \sqrt{2 G\rho_\mathrm{DM} V_\mathrm{cav}\, \omega} 
    \,,\label{eq:DM_cavity_coupling}
\end{equation}
where 
\begin{equation}\label{eq:GeometricFactor}
G\equiv \left|\frac{\int d^3x\, \hat E \cdot \vec n}{\sqrt{V_\mathrm{cav}\int d^3x\, |\hat E|^2 }}\right|^2   
\end{equation}
is the mode overlap factor for the mode $\hat E$ and $\vec n$ is the polarization unit vector of the dark matter field. When averaged over many coherence times, the polarization average in $G$ leads to a factor of 1/3, and the spatial overlap is of order~$0.1-1$ for a well-chosen mode.

Going into the rotating frame with mode $a$, we obtain
\begin{align}
    H_\textrm{int} = g \cos\left( m_{A'}t + \phi(t)\right)
     a^\dagger e^{i\omega t} +\mathrm{h.c.}\approx g e^{i(\omega-m_{A'})t+i\phi(t)} a^\dagger +\mathrm{h.c.}
\end{align}
where in the second equation we applied the rotating wave approximation and dropped the rapidly oscillating term. 

\section{Continuous Stochastic Model for Dark Photon-Cavity Interaction}
\label{app:random_walk}
We model the dark photon’s interaction with a cavity as a stochastic drive with a finite correlation time $ \tau_\text{DM} $, represented by a linear drive term $ g(t) a + g^*(t) a^\dagger $ acting on the cavity field operator $ a $. Here, $ g(t) = |g| e^{i(\Delta t+\phi(t))} $ is the linear drive strength, with $\Delta = \omega_\textrm{DM}-\omega_\textrm{cav}$ the drive detuning and and $ \phi(t) $ the DM random phase. Note that the polarization of the dark photon field also varies over the correlation time. Here we keep $\lvert g\rvert$ constant and capture this effect by including a 1/3 factor in the geometric factor, see Sec.~\ref{app:dark photon interaction}. Assuming the cavity lifetime is long compared to the integration time, we neglect damping, and the cavity field evolves under the Hamiltonian $ H = \hbar (g(t) a + g^*(t) a^\dagger) $. The equation of motion is $ \dot{a} = -i [a, H] / \hbar $, yielding:
\begin{align}
\dot{a} = -i g(t), \quad a(t) = a(0) - i \int_0^t g(t') \, dt'.
\label{eq:EOM}
\end{align}
For simplicity, assume $ a(0) = 0 $, so the displacement is $ \alpha(t) = a(t) = -i \int_0^t |g| e^{i(\Delta t'+\phi(t'))} \, dt' $, and the cavity population is $ \langle n(t) \rangle = \langle |\alpha(t)|^2 \rangle = \left| \int_0^t |g| e^{i(\Delta t'+\phi(t'))} \, dt' \right|^2 $. The expectation value is:
\begin{align}
\langle |\alpha(t)|^2 \rangle &= \langle \alpha^*(t) \alpha(t) \rangle = \int_0^t \int_0^t |g|^2 e^{i\Delta(t'-t'')}\langle e^{i\phi(t')} e^{-i\phi(t'')} \rangle \, dt' dt''.
\end{align}
Assuming $ \phi(t) $ decorrelates over $ \tau_\text{DM} $, we model the correlation as $ \langle g^*(t) g(t') \rangle = |g|^2 \langle e^{i\phi(t)} e^{-i\phi(t')} \rangle = |g|^2 e^{-|t - t'| / \tau_\text{DM}} $, reflecting exponential coherence decay. Thus:
\begin{align}
\langle n(t) \rangle &= |g|^2 \int_0^t \int_0^t e^{i\Delta(t'-t'')}e^{-|t' - t''| / \tau_\text{DM}} \, dt' dt''.
\end{align}
Using symmetry and taking the real part:
\begin{align}
\langle n(t) \rangle &= 2 |g|^2 \int_0^t dt' \int_0^{t'} dt'' \, \cos(\Delta (t' - t'')) e^{-(t' - t'') / \tau_\text{DM}}.
\end{align}
Substitute $ u = t' - t'' $, and let $ \kappa = \tau_\text{DM}^{-1} $ represent the DM linewidth, the inner integral becomes:
\begin{align}
\int_0^{t'} \cos(\Delta u) e^{-\kappa u} \, du = 
\frac{\kappa}{\kappa^2 + \Delta^2} - \frac{e^{-\kappa t'} (\kappa \cos(\Delta t') - \Delta \sin(\Delta t'))}{\kappa^2 + \Delta^2}.
\end{align}
Integrating over $ t' $, after evaluating each term and simplifying, we obtain:
\begin{align}
\langle n(t) \rangle &= \frac{2 g^2}{\kappa^2 + \Delta^2} \left[ at  + e^{-at}\frac{\left(\kappa^2 - \Delta^2\right)  \cos(\Delta t)-2 \kappa \Delta\sin(\Delta t)}{\kappa^2 + \Delta^2} - \frac{\kappa^2 - \Delta^2}{\kappa^2 + \Delta^2} \right].\label{eq:n(t)}
\end{align}

\begin{figure*}[t!]
\centering
\includegraphics[width=0.8\textwidth]{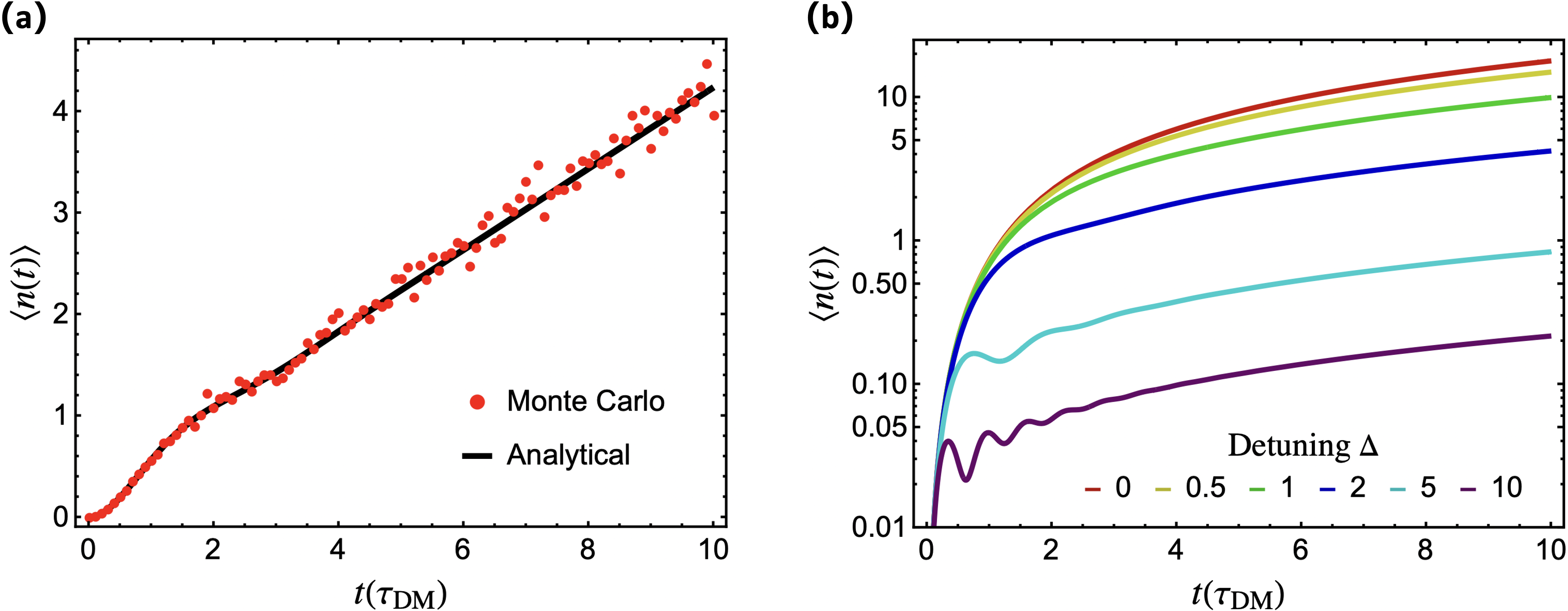}
\captionsetup[figure]{justification=justified,singlelinecheck=false}
\caption{\justifying
\textbf{Cavity population induced by DM.} (a) The analytical result of the cavity population from Eq.~\eqref{eq:n(t)} showing excellent agreement with the Monte Carlo calculation. In this simulation, we set $g=\tau_\textrm{DM}^{-1}=\Delta/2$. (b) Cavity population at different detunings. Under larger detuning, a transient oscillation can be observed in the initial time until it decays on the timescale of $\tau_\textrm{DM}$.}
\label{fig:random_walk}
\end{figure*}

In Fig.~\ref{fig:random_walk}(a), we compare this analytical formula to the numerical result obtained from the Monte Carlo method, showing excellent agreement between the two. In the latter, we numerically solve the Heisenberg equation of motion for the cavity field operator $ a $ in Eq.~\eqref{eq:EOM}. For each Monte Carlo trajectory, we generate a continuous phase trace $\phi(t)$ via phase diffusion. We draw an initial phase $\phi(0)$ uniformly from $[0,2\pi)$, and then evolve it on a fine time grid $t_k = k\,\Delta t$ with independent Gaussian increments,
\begin{equation}
\phi(t_{k+1}) = \phi(t_k) + \delta\phi_k, 
\qquad 
\delta\phi_k \sim \mathcal{N}\!\left(0,\; \frac{2\Delta t}{\tau_{\rm DM}}\right),
\end{equation}
so that the complex phasor obeys $\left\langle e^{i\left[\phi(t+\tau)-\phi(t)\right]} \right\rangle
= e^{-|\tau|/\tau_{\rm DM}}$. For each trajectory, we compute the cavity displacement $\alpha(t)$ by integrating the driven cavity response, and then ensemble-average $n(t)=|\alpha(t)|^2$ over $10^3$ trajectories. We also examine the effect of the DM-cavity detuning on the signal population at different times. As demonstrated in Fig.~\ref{fig:random_walk}(b), the detuning reduces the signal population at all times, but the reduction effect is much more pronounced for longer times than for the short time limit. Under large detuning, oscillations of the signal population can happen at the beginning of time. Since we limit ourselves to detecting dark photons within the dark photon linewidth, i.e., $\Delta\leq \kappa=\tau_\text{DM}^{-1}$, we are away from the oscillatory regime and expect the population to grow monotonically over time.

In the resonant case ($\Delta = 0$), Eq.~\eqref{eq:n(t)} reduces to 
\begin{align}
    \langle n(t)\rangle=2 g^2  \tau_\textrm{DM}\left[ t  -\tau_\textrm{DM} \left(1-e^{-t/\tau_\textrm{DM}}\right) \right],
\end{align}
which corresponds to the effective displacement amplitude in the main text Eq.~\eqref{eq:DM_displacement},  
\begin{equation}
\langle\lvert\alpha(\tau_\mathrm{int})\rvert\rangle = 
    \sqrt{2 |g|^2 \tau_\mathrm{DM} \left[ \tau_\mathrm{int} - \tau_\mathrm{DM} \left(1 - e^{-\frac{\tau_{\text{int}}}{ \tau_\mathrm{DM}}}\right) \right]}\,.
\end{equation}

\section{Entanglement-enhanced DM-induced cavity displacement}
\label{app:Entanglement-enhanced-displacement}

\begin{figure}[b]
\begin{subfigure}[]{.70\textwidth}
\includegraphics[width=\textwidth]{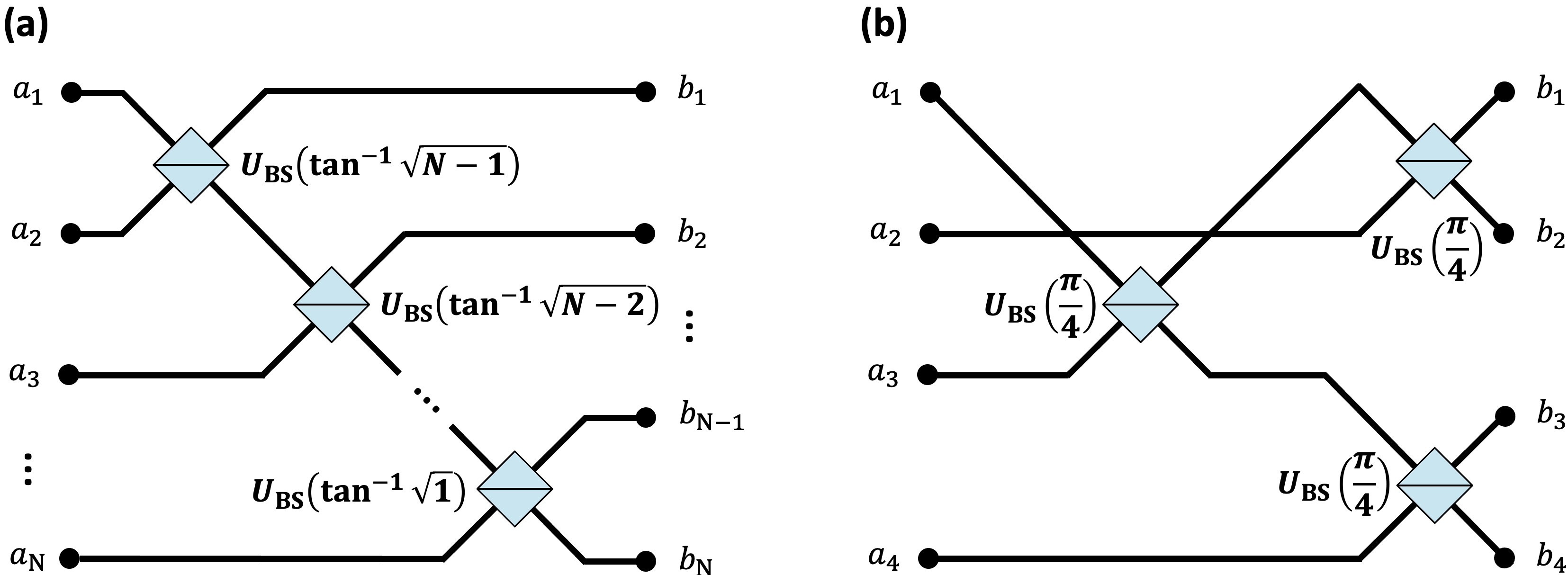}
\end{subfigure}
\caption{\justifying Quantum circuits for the entanglement distribution operation, $U_\text{ED}$, with (a) linear and (b) binary array of beamsplitters (showing only four cavities for simplicity of illustration).}
\label{fig:beamsplitters}
\end{figure}

The successful generation of the enhanced displacement of $D_1(\sqrt{N}\alpha)$ (Eq.~\eqref{eq:disp_enhancement} in the main text) is a central requirement of our protocol. Here, we prove how this is realized by the entanglement distribution operation $U_{ED}$ and the inverse operation $U_{ED}^{-1}$, which satisfy the following relation, 
\begin{align}
    U_{ED}^{-1} \sum_{n=1}^N \hat{a}^\dagger_n U_{ED} = \sqrt{N} \hat{a}^\dagger_1.\label{eq: 7}
\end{align}

\begin{align}
    U_{ED}^{-1} \sum_{n=1}^N \hat{a}_n U_{ED} = \sqrt{N} \hat{a}_1.\label{eq: 8}
\end{align}

Under this transformation, the collective displacement of the $N$ cavities due to the dark photon becomes

\begin{equation*}
U^{-1}_\text{ED}D_1(\alpha)\otimes D_2(\alpha)\cdots\otimes D_N(\alpha)U_\text{ED}= U^{-1}_\text{ED}e^{(\alpha\sum_{n=1}^N a^\dagger_n-\alpha^* \sum_{n=1}^N a_n)}U_\text{ED}=e^{\sqrt{N}(\alpha a^\dagger_1-\alpha^* a_1)}=D_1(\sqrt{N}\alpha),
\end{equation*}
giving rise to the enhanced displacement of the primary cavity.

The entanglement distribution and its inverse can be constructed by an array of microwave beamsplitters. The pairwise beamsplitter interaction between two cavities is given by
\begin{equation}
    U_{\text{BS}}(\theta,\varphi) = e^{i\theta(e^{i\varphi}a^\dagger b +e^{-i\varphi}a b^\dagger)}, \quad\begin{pmatrix}
        a\\
        b
    \end{pmatrix}\xrightarrow{U_{\text{BS}}(\theta,\varphi)}\begin{pmatrix}
\cos\theta & ie^{i\varphi}\sin\theta \\
ie^{-i\varphi}\sin\theta & \cos\theta 
\end{pmatrix}\begin{pmatrix}
        a\\
        b
    \end{pmatrix},
    \label{eq:UBS}
\end{equation}
where $a$ and $b$ are the ladder operators of the two cavities, $\theta$ and $\varphi$ are the rotation angle and the phase angle of the microwave beamsplitter, which are fully tunable leveraging circuit-QED techniques~\cite{Gao2018, lu2023high}.

The construction of $U_\text{ED}$ through beamsplitters is not unique. Two examples are illustrated in Fig.~\ref{fig:beamsplitters}, where $U_\text{ED}$ is realized through a linear sequence of beamsplitters between adjacent cavities, or through a binary sequence of 50:50 beamsplitters ($\theta =\pi/4$). For both schemes, $N-1$ beamsplitters are needed to connect all the cavities. However, the binary scheme offers a significant time advantage, scaling logarithmically with the number of cavities as 
$O(\log N)$, compared to the linear scheme’s 
$O(N)$ scaling. On the other hand, the binary scheme demands a higher connectivity than the linear scheme, introducing additional engineering and control overhead. Further optimization of the scheme can be achieved by simultaneously activating all beamsplitters in the array, implementing a global Hamiltonian that drives the system to the target state in $O(1)$ time~\cite{Christandl2004, Guery-Odelin2019}. For simplicity, we focus on the binary scheme for this work.

\section{Parasitic Effects of Dissipative Channels} 
\label{app: parasitic}
\subsection{Transformed cavity loss channels and their effect on DM integration}
\label{app:parasitic1}

In this section, we analyze the effect of excitation, decay and dephasing on an $N$-cavity sensing array during the integration time. Here, we consider perfect entanglement operations. We aim to demonstrate that the effect of thermal excitation across all cavities in our protocol is equivalent to applying an averaged excitation operator to the primary cavity, and decay and dephasing only contribute to non-heating effects, such as base population reduction. Specifically, we show how these noise channels result in a background population in $\lvert m+1\rangle
$ and a loss of base population in $\lvert m\rangle
$, at rates independent of the number of cavities $N$ but dependent on the number of Fock state photons $m$, critical to the enhancement factor of $N^2(m+1)$ in the SNR of our scheme.

Since in the following, we will extensively apply the entanglement unitary to the ladder operators, it is convenient to write the initial state of the $N$-sensor network as:

\begin{equation}
    |\psi_0\rangle = |m, 0, \ldots, 0\rangle = \frac{1}{\sqrt{m!}} (a^\dagger_1)^m |0, 0, \ldots, 0\rangle.
\end{equation}

The system undergoes an entanglement operation via the unitary $ U_{\text{ED}} $, transforming the state into a symmetric superposition where the $ m $ photons are distributed across all $ N $ cavities:

\begin{equation}
    |\psi_1\rangle = U_\text{ED}|\psi_0\rangle=\frac{1}{\sqrt{m!}} \left[ \frac{1}{\sqrt{N}} (a^\dagger_1 + a^\dagger_2 + \cdots + a^\dagger_N) \right]^m |0, 0, \ldots, 0\rangle.
\end{equation}

During the dark photon interaction phase, loss events occur in the individual cavities, transforming the entangled state $ |\psi\rangle_1 $ into a mixed state. Making use of the standard Kraus operator formalism~\cite{KrausFormalism}, we can approximate the density matrix after such an event as:

\begin{align}
    \rho &\approx \sqrt{1-\sum_{n=1}^N P_n A^\dagger_n A_n} |\psi_1\rangle \langle\psi_1| \sqrt{1-\sum_{n=1}^N P_n A_n A^\dagger_n} + \sum_{n=1}^N P_n A_n |\psi_1\rangle \langle\psi_1| A^\dagger_n \\ 
    &\approx |\psi_1\rangle \langle\psi_1| - \frac{1}{2} \sum_{n=1}^N P_n (A^\dagger_n A_n |\psi_1\rangle \langle\psi_1| + |\psi_1\rangle \langle\psi_1| A_n A^\dagger_n) + \sum_{n=1}^N P_n A_n |\psi_1\rangle \langle\psi_1| A^\dagger_n,
\end{align}
where $A_n$ represents the loss operator associated with one of the loss events (excitation $a^\dagger_n$, decay $a_n$, and dephasing $a^\dagger_na_n$). The probability for the event to happen to the $ n $-th cavity is determined by the loss rate and the integration time, $P_n = \gamma_{A, n}\tau_{int}$,
and we neglect multiple loss events by assuming $ P_n $ is small.

Finally, we apply the inverse entanglement distribution $ U_{\text{ED}}^{-1} $ to the system, resulting in the final density matrix:

\begin{equation}
\label{eq:rho_f}
    \rho_f  = U_\text{ED}^{-1}\rho U_\text{ED}=  |\psi_0\rangle \langle\psi_0| - \frac{1}{2} \sum_{n=1}^N P_n (\tilde{A}^\dagger_n \tilde{A}_n |\psi_0\rangle\langle\psi_0| + |\psi_0\rangle \langle\psi_0| \tilde{A}_n \tilde{A}^\dagger_n) + \sum_{n=1}^N P_n \tilde{A}_n |\psi_0\rangle \langle\psi_0| \tilde{A}^\dagger_n,
\end{equation}
with $\tilde{A}_n=U_\text{ED}^{-1}A_n U_\text{ED}$. This tells us that the entanglement distribution and the inverse entanglement distribution effectively transform loss channels $A$ during the integration into new channels $\tilde{A}$ that act on the initial Fock state, $|\psi\rangle_0$. From Eqs.~\eqref{eq: 7} and \eqref{eq: 8}, the transformation of the excitation ($A_n = a^\dagger_n$), decay ($A_n = a_n$) and dephasing ($A_n = a^\dagger_na_n$) processes can be written down as 
\begin{align}
U_\text{ED}^{-1}a_n^\dagger U_\text{ED} &= \frac{1}{\sqrt{N}}a_1^\dagger + \sum_{n'=2}^N c^*_{nn'}a^\dagger_{n'},\label{eq: excitation_transformed} \\
U_\text{ED}^{-1}a_n U_\text{ED} &= \frac{1}{\sqrt{N}}a_1 + \sum_{n'=2}^N c_{nn'}a_{n'},\label{eq:decay_transformed} \\
U_\text{ED}^{-1}a^\dagger_na_n U_\text{ED} &= \frac{1}{N}a^\dagger_1 a_1 + \frac{1}{\sqrt{N}}\sum_{n'=2}^N c_{nn'}a_{n'}a^\dagger_1+ \frac{1}{\sqrt{N}}\sum_{n'=2}^N c^*_{nn'}a^\dagger_{n'}a_1+\sum_{n'=2}^Nc^*_{nn'}a^\dagger_{n'}\sum_{n'=2}^N c_{nn'}a_{n'}.\label{eq:dephasing_transformed}
\end{align}
While the exact values of the coefficients $ c_{nn'} $ (for $ j \neq 1 $) depend on the specific beamsplitter construction of $ U_{\text{ED}} $, they always satisfy $\sum_{n'=2}^N \lvert c_{nn'}\rvert^2=1-\frac{1}{N}$. 

Eqs.~\eqref{eq: excitation_transformed}--\eqref{eq:dephasing_transformed} reveal the different effects of the transformed loss channels. The single-mode excitation and decay operators are transformed into collective excitation and decay of all modes, with a $1/\sqrt{N}$ coefficient for the primary mode. Summing up the contribution from all $N$-cavity loss channels, and assuming they are independent of each other, the transformed excitation deposits thermal population in $\lvert m+1\rangle$ of the primary cavity at a rate of
\begin{align}
    \gamma_{m+1} = \frac{m+1}{N}\sum_{n=1}^{N}\gamma_{\uparrow,n} = (m+1)\bar{\gamma}_{\uparrow 1},\label{eq:gamma_m+1}
\end{align}
while the transformed decay depletes the base population of primary cavity $\lvert m\rangle$
at a rate of 
\begin{align}
    \gamma_{m} = \frac{m}{N}\sum_{n=1}^{N}\gamma_{\downarrow,n} = m\bar{\gamma}_\downarrow.\label{eq:gamma_m}
\end{align}
Here, $\gamma_{\downarrow(\uparrow),i}$ represents the decay (heating) rate of the i-th cavity,
and $\bar{\gamma}_{\downarrow(\uparrow)}$ is the mean of the cavity decay(heating) rates. 

The dephasing operator, on the other hand, transforms to a more complicated combination of dephasing and photon exchange terms. Importantly, the third term on the right-hand-side of Eq.~\eqref{eq:dephasing_transformed} represents photon swapping from the primary cavity to other cavities, resulting in an effective reduction of the primary cavity $\lvert m\rangle$ population at a rate of 
\begin{align}
    \gamma_{m}^\phi = \frac{m}{N}\sum_{n=1}^{N}\gamma_{\phi,n}(\sum_{n'=2}^N\lvert c_{nn'}\rvert^2) = m(1-\frac{1}{N})\bar{\gamma}_\phi,\label{eq:gamma_phi}
\end{align}
with $\gamma_{\phi,i}$ and $\bar{\gamma}_\phi$ representing the individual and the mean of cavity dephasing rates.

These transformed loss channels can significantly affect the performance of the DM sensing protocol. Importantly, as the primary source of false positives in this scheme, the transformed excitation is independent of the number of cavities, critical to preserving the enhancement factor of $N^2(m+1)$ for the scan rate, one of the central results in this work. Meanwhile, both decay and dephasing result in effective photon loss in the primary cavity, causing a reduction in the sensing efficiency and ultimately leading to the loss of SNR in the long-term limit. Similar to the transformed excitation, the transformed decay also has no dependence on the number of cavities, whereas the transformed dephasing only has weak dependence that saturates at large cavity numbers.

\subsection{Simulating the SNR in the presence of integration and beamsplitter loss}\label{app:simulation}
In this section, we numerically analyze the effect of the noise channels on the signal-to-noise ratio (SNR). We define SNR as 
\begin{align}
    \textrm{SNR}(\tau_\textrm{int}) =\frac{n_s(\tau_\textrm{int})*n_\textrm{cycle}}{\sqrt{n_b(\tau_\textrm{int})*n_\textrm{cycle}}}=\frac{n_s(\tau_\textrm{int})*\frac{\tau_\textrm{tot}}{\tau_\textrm{cycle}}}{\sqrt{n_b(\tau_\textrm{int})*\frac{\tau_\textrm{tot}}{\tau_\textrm{cycle}}}}= \frac{\mathcal{R}_s(\tau_\textrm{int})}{\sqrt{\mathcal{R}_b(\tau_\textrm{int})}}\sqrt{\tau_\textrm{tot}}.\label{eq:app_snr}
\end{align}
Here, $\tau_\textrm{int}$ is the DM integration time in a single cycle, $\tau_\textrm{cycle} = \tau_\textrm{int} + 2\tau_\textrm{ED}+\tau_\textrm{SPAM}$ is the time for a single cycle, including the integration time, the entanglement and inverse entanlgement time, and the state preparation and measurement (SPAM) time. The total time cost of the sensing experiment, $\tau_\textrm{tot}$, consists of $n_\textrm{cycle}\gg1$ cycles of measurements.

Both $n_s$ and $n_b$ will be affected by the various loss channels in the system. To precisely capture their effects, we employ the Discretized Lindblad Master Equation (DLME) approach for numerically solving the evolution of the open system. We separately calculate the signal population $n_s$ and the background population $n_b$, assuming no heating error for the former and no signal strength for the latter,

\begin{align}
    \rho(t+\Delta t) &= U_\textrm{disp}[\Delta\alpha (t)]\rho(t) U_\textrm{disp}^{-1}[\Delta\alpha (t)] + \Delta t \sum_{n=1}^N \left(A_n\rho(t)A_n^\dagger-\frac{1}{2}\left\{A_n^\dagger A_n,\rho(t)\right\}\right),\label{eq:dlme} \\
n_s(t) &= \langle \psi_0\lvert \rho(t) \rvert \psi_0 \rangle \Big|_{T=0}\,, \quad n_b(t) = \langle \psi_0\lvert \rho(t) \rvert \psi_0 \rangle\Big|_{g(t)=0} \,. \label{eq:populations}
\end{align}
Here, $U_\textrm{disp}[\Delta\alpha (t)]$ is the unitary implementing the incremental displacement $\Delta\alpha (t)$, that represents the DM-induced cavity displacement, as
\begin{align}
    U_\textrm{disp}[\Delta\alpha (t)] &= e^{\Delta\alpha (t)\sum_{n=1}^N a^\dagger_n-\Delta\alpha(t)^* \sum_{n=1}^N a_n}, \\
    \Delta\alpha (t) & = \sqrt{2 g^2 \tau_\textrm{DM}\left[ 1  -\frac{\tau_\textrm{DM}}{t+\Delta t} \left(1-e^{-\frac{t+\Delta t}{\tau_\textrm{DM}}}\right) \right]}-\sqrt{2 g^2 \tau_\textrm{DM}\left[ 1  -\frac{\tau_\textrm{DM}}{t} \left(1-e^{-\frac{t}{\tau_\textrm{DM}}}\right) \right]}.\label{eq:deltaalphat}
\end{align}

\begin{figure}[t]
\centering
\includegraphics[width=1
\textwidth]{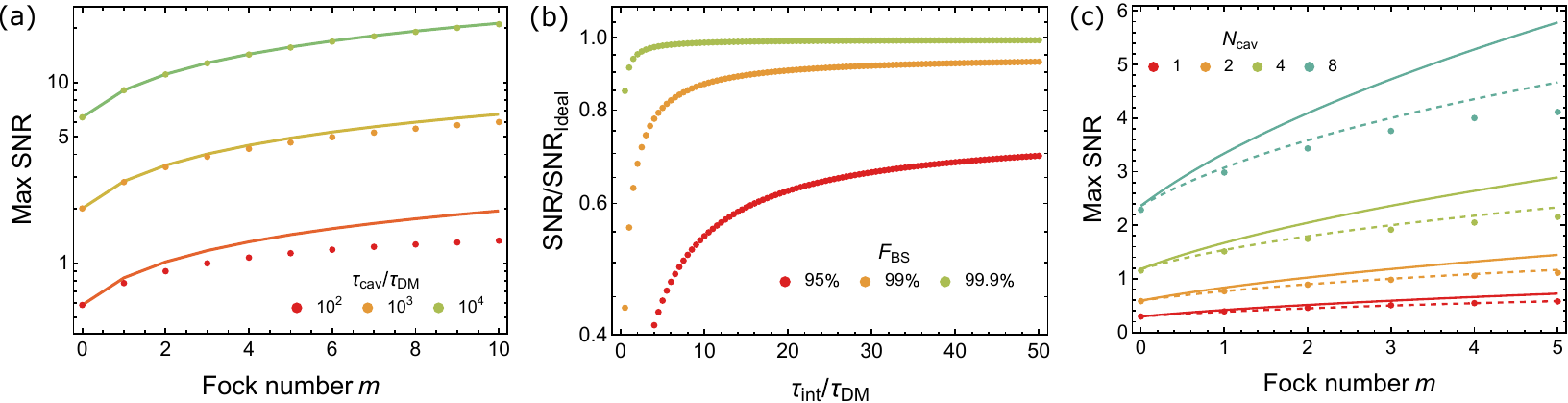}
\caption{\justifying
\textbf{The effects of cavity decoherence and beamsplitter infidelity on the detection SNR.} (a) The maximum SNR (dots) found from all integration times as a function of Fock number, showing good consistency with the expected scaling of $\sqrt{m+1}$ (solid lines). (b) SNR as a function of the beamsplitter fidelity, in units of ideal SNR in the absence of loss. (c) Comprehensive result showing maximum SNR in the presence of both integration loss and beamsplitter infidelity. Solid lines are the ideal scaling of $N\sqrt{m+1}$. Dashed lines are from fitting the $N=1$ result (red dots) and scaling it by multiplying the cavity number, $N$.}
\label{fig:SNR_vs_m}
\end{figure}

Using Eqs.~\eqref{eq:app_snr}--\eqref{eq:deltaalphat}, we calculate the SNR subject to loss during integration, assuming perfect beamsplitter operations (i.e. perfect initial state, $\rho(0) = \lvert\psi_1\rangle\langle\psi_1\rvert$). For concreteness, we assume a two-cavity sensor array at \SI{7}{\giga\hertz} at a thermal temperature of \SI{50}{\milli\kelvin}. Since nonlinear Josephson devices are needed for the entanglement operations, we further assume they introduce a pure cavity dephasing rate of 1/10 of the cavity decay rate, even during idling (integration). We set the DM interaction strength to be $g = 2\pi\times \SI{11.7}{\hertz}$, and sweep the integration time $\tau_\textrm{int}$ from 0 to 20$\tau_\textrm{DM}$. In each sweep,  we extract the maximum SNR from all integration times, and plot them as a function of Fock number $m$, for different cavity quality factors (see Fig.~\ref{fig:SNR_vs_m}(a)). The numerical result is in good agreement with the ideal scaling of $\sqrt{m+1}$, shown as solid lines. The deviation at large Fock number of lower cavity quality factor can be intuitively understood as an elevated depletion of the $\lvert m+1\rangle$ population, preventing the sufficient accumulation of signal that would require an integration up to a few $\tau_\textrm{DM}$. 

In addition to parasitic processes occurring during the integration time, the entanglement and disentanglement distribution, composed of beamsplitter operations, can also introduce errors due to the beamsplitter infidelity. In the domain of circuit QED with microwave cavities, the beamsplitters are typically implemented by stimulating Josephson nonlinearity that mediates parametric conversion processes between the cavity modes~\cite{Gao2018,chapman2023,lu2023high,Maiti2025}. The strong pump tones needed for activating the beamsplitter interaction can also lead to parasitic effects, such as drive-induced decay and dephasing, which can significantly reduce the beamsplitter fidelity~\cite{Yaxing2019}. Recent advances in circuit QED have enabled fast beamsplitter operations ($\tau_\textrm{BS}\sim 100\textrm{ns}$) with fidelities as high as $\mathcal{F}_{\textrm{BS}}\sim$99.98\%, between 3D superconducting cavities~\cite{lu2023high}. Here, we investigate the impact of beamsplitter infidelity on our sensing protocol. For simplicity, without loss of generality, we model this infidelity as an elevated decoherence rate of the cavity modes, mimicking drive-induced decoherence effects. In a two-cavity sensor array, we set the baseline decay rate of the sensor to be $\tau_\textrm{DM}^{-1}$/100, the baseline heating rate to correspond to a \SI{50}{\milli\kelvin} thermal temperature, and the baseline dephasing rate to 1/10 of the cavity decay rate. Assuming a beamsplitter rate of $g_\textrm{BS} = 2\pi\times \SI{1}{\mega\hertz}$, we evenly increase the three loss rates of the two cavities connected by the beamsplitter, to match the set beamsplitter fidelity. Note that the set fidelity is defined for the single-photon subspace. For higher Fock state entanglement, the effective fidelity is even lower due to the elevated decoherence rate. With the five-photon initial Fock state, we calculate the SNR to ideal (no beamsplitter loss) SNR ratio, as a function of $\tau_\textrm{int}$ under different beamsplitter fidelities in Fig.~\ref{fig:SNR_vs_m}(b). We assume no loss events (except for heating) during the integration, such that the result only reveals the effect of the beamsplitter loss. The calculation was made using Eq.~\eqref{eq:populations} for both the lossy beamsplitters and the lossless integration period. As expected, the SNR reduces as the beamsplitter infidelity grows. Since beamsplitter loss affects the system only before and after the integration, a longer integration time dilutes the beamsplitter loss effect, leading to a monotonic increase of SNR. For a beamsplitter fidelity of 99\%, which is well within the capabilities of current circuit QED platforms, the ideal SNR is reasonably preserved.

Finally, we combine the effects of both the integration loss and the beamsplitter loss. We numerically computed SNR from Eq.~\eqref{eq:populations} under both losses, assuming the same aforementioned baseline decoherence numbers, and a beamsplitter fidelity of 99.9\%. For each combination of cavity number $N\in\{1,2,4,8\}$ and Fock number $m\in\{0,1,2,3,4,5\}$, we extract the maximum SNR from different integration times, which are plotted in Fig.~\ref{fig:SNR_vs_m}(c). Large cavity number $N$ clearly improves the SNR, while at large Fock number $m$ there is a noticeable fall off from the ideal scaling $N\sqrt{m+1}$. To understand the source of this deviation, we fit the $N=1$ result by the red dashed line, which is multiplied by $N$ to generate the other dashed lines. These dashed lines are in much closer agreement with the numerical results, indicating that the deviation is mainly due to the cavity loss amplified by the stimulated emission, whereas the enhancement factor of $N$ due to entanglement is relatively well preserved.

\section{Simultaneous Noise measurement}
\label{app:simultaneous_noise_measurement}
Achieving a strong limit from the precise (low variance) determination of the background rate in Eq.~\eqref{eq:BG-rate} is vital for a sensitive search. 
This can be done by careful measurement of the heating rates in cavities 1 through $N$, say, in another nearby frequency. However, our protocol provides an opportunity to measure the effective heating rate $\gamma_{\uparrow1}$ in the signal cavity in situ by measuring the other nodes of the cavity network. To see this, we point out that the unitarity of the $U_\mathrm{ED}^\dagger$ operation is related to the total heating rates before and after ED, 
\begin{equation}
   \sum_{i=1}^N  \gamma_{\uparrow i}=\sum_{i=1}^N \bar \gamma_{\uparrow i}\,.
\end{equation} 
Combining this with Eq.~\eqref{eq:gamma-bar1} gives,
\begin{equation}
\label{eq:calib}
    \bar\gamma_{\uparrow 1} = \frac{1}{N-1}\sum_{i=2}^N \bar \gamma_{\uparrow i}\,.
\end{equation} 
Namely, the heating rate in the signal cavity in our protocol is equal to the average heating rate in the rest of the network and can thus be measured simultaneously, without interfering with the signal readout. Furthermore, for particular choices of $U_\mathrm{ED}$, the false positive rate can be equal to the average in just a subset of the other cavities. For example, in the choice made in~\cite{shu2024eliminatingincoherentnoisecoherent}, the incoherent false positive rate 
satisfies~$\bar\gamma_{\uparrow 1} =\bar\gamma_{\uparrow 2}$.

This in-principle background measurement is a feature of multi-node networks, only possible for two or more cavities. DM interacts with a specific linear combination, referred to as a signal mode. Incoherent noise is distributed evenly among the non-signal modes in the system, and a sum rule exists relating false positives to heating throughout the system. In reference~\cite{shu2024eliminatingincoherentnoisecoherent}, this type of measurement of incoherent noise was called an `elimination' of the noise, subtracting the right-hand side of Eq.~\eqref{eq:calib}, or $\bar\gamma_{\uparrow2}$ for the special choice mentioned above, from the count in the signal mode. We note, however, that these relations hold only on average. The left- and right-hand sides of Eq.~\eqref{eq:calib} suffer from independent statistical fluctuations. As a result, we view this technique as a clever method to characterize the system, but not as a new `background-free' measurement, at least in the context of the noise channels we consider.

In addition, we note that though incoherent heating and loss satisfy Eq.~\eqref{eq:calib}, dephasing noise will lead to a violation of this relation when the initial state in the primary signal cavity is in a non-vacuum Fock state. This can be seen in the third term of Eq.~\eqref{eq:dephasing_transformed}, in which dephasing noise can effectively transfer a photon from the primary cavity to one of the spectators. The converse process, which is represented by the second term of this equation, does not occur at this order because the spectator cavities are empty. 

\section{Optimal Choice of Integration Time for Dark Photon Detection}
\label{app:optimal_integration_time}

In this section, we provide a semi-classical analysis on the optimal integration time $ \tau_{\text{int}} $ for maximizing the SNR and hence the scan rate, as shown in Fig.~\ref{fig:SNR_vs_m_main} in the main text.

Intuitively, there’s an optimal $ \tau_{\text{int}} $ because of competing effects. If $ \tau_{\text{int}} $ is too short, the signal per cycle doesn’t have enough time to build up, especially since the dark photon signal requires $ \tau_{\text{int}} \gtrsim 5 \tau_\text{DM} $ to reach the diffusive regime where the signal rate saturates. Additionally, since each cycle includes overhead time ($ 2 \tau_{\text{ED}} + \tau_{\text{SPAM}} $) where no signal is collected, a short $ \tau_{\text{int}} $ means a smaller effective signal rate within $ \tau_{\text{tot}} $. On the other hand, if $ \tau_{\text{int}} $ is too long, the signal will be completely depleted by the cavity decoherence over time. 

Here we derive an approximation for the optimal $ \tau_{\text{int}}$ that balances these factors. Consider the primary cavity initially in state $ |m\rangle $. As a result of our protocol, the dark photon effectively displaces the primary cavity and induces a transition from $ |m\rangle $ to $ |m+1\rangle $ at an instantaneous rate of $\mathcal{R}_{s,\textrm{inst}}^0$, 
\begin{equation}
\mathcal{R}_{s,\text{inst}}^0(t) =\frac{d}{dt}n_{s}^0(t)=2 g^2 \tau_{\text{DM}} \left(1 - e^{-\frac{t}{\tau_{\text{DM}}}}\right) N (m + 1),\label{eq:rs0_inst}
\end{equation}
where $n_{s}^0(t)$ is the enhanced $(m+1)$-th state population, introduced in Eq.~\eqref{eq:signal_population} in the main text. The same displacement also induces a transition from $ |m\rangle $ to $ |m-1\rangle $ at instantaneous rate $\frac{m}{m+1}\mathcal{R}_{s,\textrm{inst}}^0$.
In the meantime, the collective decay and dephasing in all cavities cause transitions from $ |m\rangle $ to $ |m-1\rangle $ at rate $ m\bar{\gamma}_\downarrow + m\left(1-N^{-1}\right)\bar{\gamma}_\phi$, and from $ |m+1\rangle $ to $ |m\rangle $ at rate $ (m+1)\bar{\gamma}_\downarrow + (m+1)\left(1-N^{-1}\right)\bar{\gamma}_\phi$, see Eqs.~\eqref{eq:gamma_m} and \eqref{eq:gamma_phi} in Sec.~\ref{app: parasitic}. Denoting $\gamma_{\downarrow\textrm{eff}} = \bar{\gamma}_\downarrow+\left(1-N^{-1}\right)\bar{\gamma}_\phi$, the rate equations (without considering false positives, i.e. cavity heating) for the populations $ n_m(t) $ and $ n_{m+1}(t) $ are therefore:

\begin{align}
&\frac{d n_m(t)}{dt} = -\left[m\gamma_{\downarrow\textrm{eff}}+\frac{2m+1}{m+1}R_{s,\text{inst}}^0(t)\right] n_m(t) +  (m+1)\gamma_{\downarrow\textrm{eff}} n_{m+1}(t),\\
&\frac{d n_{m+1}(t)}{dt} = R_{s,\textrm{inst}}^0(t) n_m(t) - (m+1)\gamma_{\downarrow\textrm{eff}} n_{m+1}(t),
\end{align}
with initial conditions $ n_m(0) = 1 $, $ n_{m+1}(0) = 0 $. We treat $ R_{s,\text{inst}}^0(t) n_m(t) $ as a source term, and formally solve the rate equation for $ n_{m+1}(t) $:

\begin{equation}
\label{eq:pm+1t}
n_{m+1}(t) = \int_0^t R_{s,\text{inst}}^0(t') n_m(t') e^{-(m+1)\gamma_{\downarrow\textrm{eff}}  (t - t')} dt'.
\end{equation}
Since the DM-induced transition rate is much weaker than the cavity decay rate, we can approximate $ n_m(t) $ as 
\begin{equation}
n_m(t) \approx e^{-m\gamma_{\downarrow\textrm{eff}} t}.\label{eq:pmt}     
\end{equation}
Plugging Eqs.~\eqref{eq:pmt} and \eqref{eq:rs0_inst} into Eq.~\eqref{eq:pm+1t}, we obtain the signal rate:

\begin{equation}
\mathcal{R}_s(t) = \frac{n_{m+1}(t)}{t} \approx 2 g^2 \tau_{\text{DM}} N (m + 1) e^{-(m+1)\gamma_{\downarrow\textrm{eff}} t} \left[ \frac{ e^{\gamma_{\downarrow\textrm{eff}} t} - 1 }{\gamma_{\downarrow\textrm{eff}}t}  - \frac{ e^{\left( \gamma_{\downarrow\textrm{eff}} - \tau_{\text{DM}}^{-1} \right) t} - 1 }{(\gamma_{\downarrow\textrm{eff}} - \tau_{\text{DM}}^{-1})t} \right].\label{eq:rst}
\end{equation}

\begin{figure*}[t!]
\centering
\includegraphics[width=0.9
\textwidth]{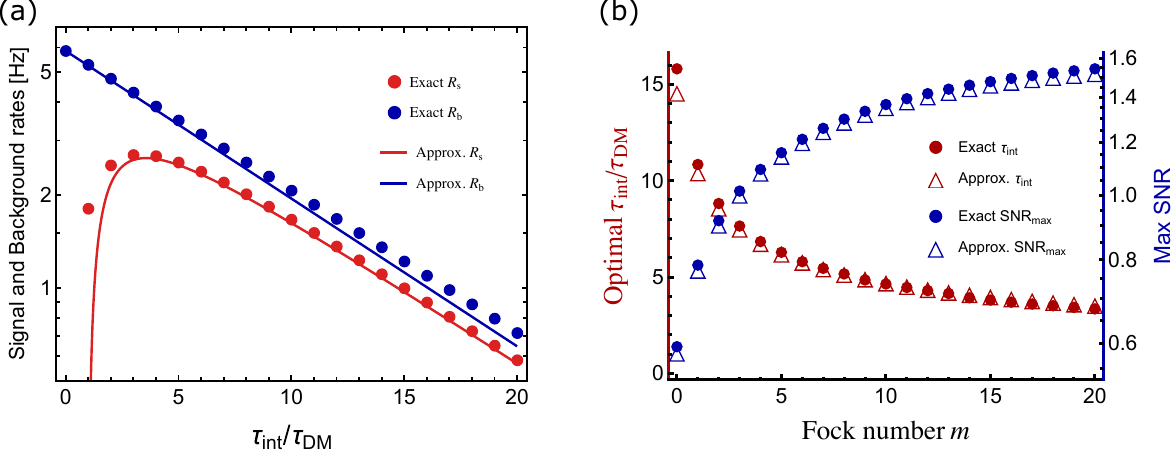}
\caption{\justifying
\textbf{Semi-classical approximation of the signal rate, background rate, optimal integration time, and maximum SNR, for the regime $\tau_\textrm{DM}<\tau_\textrm{int}\ll\gamma_{\downarrow\textrm{eff}}^{-1}$} (a) The approximated signal rate (red dots) and background rate (blue dots) for $m=10$, vs the numerical results (solid lines) from solving the Lindbladian Master Equations. (b) For each Fock number $m$, the optimal integration time and maximum SNR extracted from (a), showing close agreement between the approximation (triangles) and the exact result (dots).)}
\label{fig:snr_approximation}
\end{figure*}

On the other hand, in the absence of the DM signal but including cavity heating, the rate equations are slightly modified,
\begin{align}
&\frac{d \tilde{n}_m(t)}{dt} \approx -m\gamma_{\downarrow\textrm{eff}} \tilde{n}_m(t),\\
&\frac{d \tilde{n}_{m+1}(t)}{dt} \approx (m+1)\gamma_{\uparrow\textrm{eff}} \tilde{n}_m(t) - (m+1)\gamma_{\downarrow\textrm{eff}} \tilde{n}_{m+1}(t),
\end{align}
with $\gamma_{\uparrow\textrm{eff}}=\bar{\gamma}_\uparrow$, see Eq.~\eqref{eq:gamma_m+1}. Here, we have made approximations that the heating rate is negligible compared to the decay rate, and the second-order processes (e.g, heating followed by decay event) can be ignored.
This leads to the background rate of
\begin{align}
    \label{eq:rbt}
\mathcal{R}_b(t)
=\frac{\tilde{n}_{m+1}(t)}{t} 
=\frac{1}{t} \int_0^t (m+1)\gamma_{\uparrow\textrm{eff}} \tilde{n}_m(t') e^{-(m+1)\gamma_{\downarrow\textrm{eff}} (t - t')} dt'
\approx (m+1)\gamma_{\uparrow\textrm{eff}}e^{-(m+1)\gamma_{\downarrow\textrm{eff}} t}\frac{e^{\gamma_{\downarrow\textrm{eff}} t}-1}{\gamma_{\downarrow\textrm{eff}}t}.
\end{align}

In this analysis, we focus on high-Q cavity and the diffusive regime where $\tau_\textrm{DM}<\tau_\textrm{int}\ll\gamma_{\downarrow\textrm{eff}}^{-1}$ holds true. Under this condition, $e^{\gamma_{\downarrow\textrm{eff}} t}-1\approx \gamma_{\downarrow\textrm{eff}} t$, so we may approximate Eq.~\eqref{eq:rst} and Eq.~\eqref{eq:rbt} as
\begin{align}
\mathcal{R}_s(\tau_\textrm{int}) &\approx 2 g^2 \tau_{\text{DM}} N (m + 1) e^{-(m+1)\gamma_{\downarrow\textrm{eff}} \tau_\textrm{int}} \left( 1  - \frac{\tau_{\text{DM}}}{\tau_\textrm{int}} \right), \quad
\mathcal{R}_b(\tau_\textrm{int})
\approx (m+1)\gamma_{\uparrow\textrm{eff}}e^{-(m+1)\gamma_{\downarrow\textrm{eff}} \tau_\textrm{int}}.\label{eq:rsrb_approx}
\end{align}
Now we can rewrite the SNR in Eq.~\eqref{eq:app_snr}, as
\begin{align}
    \textrm{SNR}(\tau_\textrm{int}) = \frac{\mathcal{R}_s(\tau_\textrm{int})}{\sqrt{\mathcal{R}_b(\tau_\textrm{int})}}\sqrt{\tau_\textrm{tot}}\approx2 g^2 \tau_{\text{DM}}\sqrt{\frac{\tau_\textrm{tot}}{\gamma_{\uparrow\textrm{eff}}}} N \sqrt{m + 1} e^{-\frac{m+1}{2}\gamma_{\downarrow\textrm{eff}} \tau_\textrm{int}} \left( 1  - \frac{\tau_{\text{DM}}}{\tau_\textrm{int}} \right).\label{eq:tau_int_approx}
\end{align}
Intuitively, the factor $1-\tau_{\mathrm{DM}}/\tau_{\mathrm{int}}$ captures the slow, diffusive build-up once the DM phase decorrelates, whereas the exponential term is the survival of the $|m+1\rangle$ state under an effective decay rate $(m+1)\gamma_{\downarrow\mathrm{eff}}$. The competition between these two trends sets the optimal integration time that maximizes SNR, 
\begin{align}
    \tau_\textrm{int}^\textrm{opt}\approx \frac{\tau_\textrm{DM}}{2}\left(1+\sqrt{1+\left(\tau_\textrm{DM}\gamma_{\downarrow\textrm{eff}}(m+1)/8\right)^{-1}}\right),\quad \textrm{SNR}_\textrm{max} = \textrm{SNR}(\tau_\textrm{int}^\textrm{opt}).
\end{align}
The validity of this result for a two-cavity array is depicted in Fig.~\ref{fig:snr_approximation}, where the optimal integration times and maximum SNRs computed from Eq.~\eqref{eq:tau_int_approx} show good agreement with numerical results from the DLME (Eq.~\eqref{eq:dlme}). 

\section{Global optimum of the integration time and Fock number}
\label{app:global_opt_tau_m}
In Appendix~\ref{app:optimal_integration_time}, we optimized the integration time $\tau_{\mathrm{int}}$ at fixed target Fock number $m$. Here we extend that analysis by allowing both $\tau_{\mathrm{int}}$ and $m$ to vary, and we derive a simple high-$Q$ asymptotic estimate for the \emph{global} optimum. Importantly, the closed-form optimum obtained in Appendix~F from \eqref{eq:tau_int_approx} relies on the diffusive-regime assumption $\tau_{\mathrm{DM}} \ll \tau_{\mathrm{int}} \ll \gamma_{\downarrow\mathrm{eff}}^{-1}$. When searching for the global optimum, however, the optimal $\tau_{\mathrm{int}}$ generally decreases with increasing $m$ due to the survival factor $e^{-(m+1)\gamma_{\downarrow\mathrm{eff}}\tau_{\mathrm{int}}/2}$, and in the large-$m$ regime one can have $\tau_{\mathrm{int}}\sim O(\tau_{\mathrm{DM}})$ so that the assumption $\tau_{\mathrm{int}}\gg \tau_{\mathrm{DM}}$ is no longer controlled. We therefore start from the SNR definition \eqref{eq:app_snr} and use the semi-classical signal and background rates $\mathcal{R}_s(t)$ and $\mathcal{R}_b(t)$ derived in Appendix~\ref{app:optimal_integration_time} without invoking $\tau_{\mathrm{int}}\gg \tau_{\mathrm{DM}}$. Dropping overall prefactors independent of $\tau_{\mathrm{int}}$ and $m$, we define a dimensionless SNR proxy $S$,
\begin{equation}
S\propto \sqrt{m+1}\,e^{-(m+1)y/2}\,
\frac{f_1(y)-f_2(y)}{\sqrt{f_1(y)}}.
\label{eq:appG_Sdimless}
\end{equation}
Here we introduced 
\begin{equation}
f_1(y)\equiv \frac{e^{y}-1}{y},\qquad
f_2(y)\equiv \frac{e^{(1-\rho)y}-1}{(1-\rho)y},
\label{eq:appG_f1f2}
\end{equation}
with dimensionless variables
$y=\gamma_{\downarrow\mathrm{eff}}t$, $
\rho
=(\gamma_{\downarrow\mathrm{eff}}\tau_{\mathrm{DM}})^{-1}$. In the high-$Q$-cavity limit $\rho\gg 1$, the optimum occurs at $y\ll 1$ while the DM-scaled integration time $a\equiv t/\tau_{\mathrm{DM}}=\rho y$
remains of order unity. In this regime we can approximate $f_{1,2}(y)$ as,
\begin{equation}
f_1(y)\approx 1,
\qquad
f_2(y)\approx\frac{1-e^{-a}}{a}.
\label{eq:appG_f_asympt}
\end{equation}
Substituting Eq.~\eqref{eq:appG_f_asympt} into Eq.~\eqref{eq:appG_Sdimless} gives
\begin{equation}
S\propto \sqrt{b\rho}\,
\exp\!\left[-\frac{ab}{2}\right]\,
h(a),
\quad
h(a)\equiv \frac{a-1+e^{-a}}{a},\quad b\equiv\frac{m+1}{\rho},
\label{eq:appG_S_asympt}
\end{equation}
where $b$ parametrizes the target Fock number measured in units of the large dimensionless ratio $\rho=(\gamma_{\downarrow\mathrm{eff}}\tau_{\mathrm{DM}})^{-1}$. We first find out the optimal $b$ (denoted as $b^*$) for maximizing $S$ under fixed $a$, i.e.,
\begin{equation}
\frac{\partial}{\partial b}\ln S=\frac{1}{2b}-\frac{a}{2}=0
\quad\Rightarrow\quad
b^\ast=\frac{1}{a}.
\label{eq:appG_b_opt}
\end{equation}
Substituting Eq.~\eqref{eq:appG_b_opt} back into Eq.~\eqref{eq:appG_S_asympt} and optimizing over $a$ yields a single scalar equation,
\begin{equation}
\frac{\partial}{\partial a}\ln S=-\frac{1}{2a}+\frac{d}{da}\ln{h(a)}=0,
\label{eq:appG_a_stationary}
\end{equation}
with a numerical solution of $a^\ast \simeq 2.15$.
Therefore, in the high-$Q$ limit the globally optimal integration time scales as
\begin{equation}
(\tau_{\mathrm{int}}^{\mathrm{opt}})^*
\simeq 2.15\tau_{\mathrm{DM}},
\label{eq:appG_tau_opt}
\end{equation}
with corresponding global optimal Fock number
\begin{equation}
(m_{\mathrm{opt}})^*+1=b^\ast\rho=\frac{\rho}{a^\ast}
\simeq \frac{0.465}{\gamma_{\downarrow\mathrm{eff}}\tau_{\mathrm{DM}}}.
\label{eq:appG_m_opt}
\end{equation}
In practice, $m$ is an integer, so \eqref{eq:appG_m_opt} should be rounded to the nearest integer.

\section{Effects of SPAM on SNR}
\label{app:other_background}

So far in our analysis, we have focused on the entanglement and integration sections of our detection protocol. In this appendix section, we examine the effect of state preparation and measurement (SPAM) on the SNR, and discuss mitigation techniques. 

\noindent{\bfseries Initial state $\lvert m\rangle$ preparation.}
Nonlinear ancillary devices such as transmon qubits allow universal control of the cavity modes, including the preparation of Fock states~\cite{SNAP, ECD, Deng2024,Huang2025,Kim2025}. The time cost and fidelity of preparation are device and protocol-dependent. In the recent example of the sideband feedforward control scheme~\cite{Kim2025}, Fock states up to twenty photons can be prepared with $>95\%$ fidelity in $<$\SI{10}{\micro\second}, with most of the infidelity due to leakage into lower Fock states. In addition, repeated measurement and post-selection on NOT $\lvert m+1\rangle$ right after the preparation of $\lvert m\rangle$ further suppresses false positives due to preparation error.

\noindent{\bfseries Signal state $\lvert m+1\rangle$ measurement.} After the inverse entanglement step, a cavity $\lvert m+1\rangle$ state measurement is crucial for finding the DM signal. The cavity Fock state can be mapped to the ancillary qubit via a conditional $\pi$-pulse. However, qubit thermal population and readout assignment errors can both affect the performance of this measurement. Suppose the qubit's population in the excited state $ |e\rangle $ is $ P_e $ at thermal equilibrium, and the ground state population $ P_g = 1 - P_e $. The qubit-cavity system state, after the inverse entanglement distribution, is a product state reflecting the thermal mixture of the qubit and the primary cavity:
\[
\rho_{\text{qubit-cavity}} \approx \left[ P_g |g\rangle\langle g| + P_e |e\rangle\langle e| \right] \otimes \left[ (1-p_{m+1}) |\overline{m+1}\rangle\langle \overline{m+1}| + p_{m+1} |m+1\rangle\langle m+1| \right].
\]
Here, $|\overline{m+1}\rangle$ represents all primary cavity states other than $|m+1\rangle$. For concreteness, let's consider the conditional $\pi$-pulse scheme~\cite{PhysRevLett.132.140801, Kim2025}. The conditional operation ensures that if the cavity is in $ |m+1\rangle $, the qubit is flipped; otherwise, the qubit remains unchanged. Then, a measurement outcome of $\lvert e\rangle$ would be treated as a positive event. The measured $ |e\rangle $-state probability per cycle, accounting for the qubit’s thermal population and readout error, is:
\begin{align}
P(|e\rangle) = \left[P_e (1-\mathcal{E}_{eg}) +P_g\mathcal{E}_{ge}\right](1-p_{m+1})  + \left[P_g(1-\mathcal{E}_{eg})+P_e\mathcal{E}_{ge}\right] p_{m+1} .
\end{align}
Here, $\mathcal{E}_{ij}$ represents the probability of finding the state in $\lvert j\rangle$ for an input state of $\lvert i\rangle$. The effective measured background rate $ \mathcal{R}_b' $ per cycle (without signal, i.e., $ p_{m+1} \approx \mathcal{R}_b \tau_{\text{cycle}} $) is:
\begin{align}
\mathcal{R}_b' = \frac{\left[P_e (1-\mathcal{E}_{eg}) +P_g\mathcal{E}_{ge}\right] (1 - \mathcal{R}_b \tau_{\text{cycle}})  +\left[P_g(1-\mathcal{E}_{eg})+P_e\mathcal{E}_{ge}\right]  (\mathcal{R}_b \tau_{\text{cycle}})}{\tau_{\text{cycle}}}\approx (P_e+\mathcal{E}_{ge}) \tau_\textrm{cycle}^{-1}+\mathcal{R}_b,
\end{align}
assuming $P_g\approx 1$, $\mathcal{R}_b \tau_\textrm{cycle}\approx 0$, $P_e\mathcal{E}_{eg}\approx P_e\mathcal{E}_{ge}\approx 0$. While it appears that the qubit thermal population $P_e$ and readout error $\mathcal{E}_{ge}$ would increase the background rate, they can be effectively eliminated by performing repeated cavity measurements, and resetting the qubit state in between measurements. After repeating $n$ times, the background rate becomes
\begin{align}
\mathcal{R}_b' \approx (P_e^n+\mathcal{E}_{ge}^n) \tau_\textrm{cycle}^{-1}+\mathcal{R}_b,
\end{align}
which approaches the original background rate $\mathcal{R}_b$ under sufficient $n$. 

\noindent{\bfseries Cavity reset.} 
For high-Q cavities, the relaxation of the large photon Fock state back to the vacuum state could become the dominant time cost in each cycle of detection. Since $\textrm{SNR}\propto\tau_\textrm{cycle}^{-1/2}$ (under fixed $n_s$ and $n_b$, see Eq.~\eqref{eq:app_snr}), the long duty cycle could severely limit the SNR of detection. This can be again circumvented by leveraging the qubit-cavity interaction for engineering cavity reset operations, providing speedups orders of magnitude faster than the natural cavity decay rate~\cite{Huang2025,Kim2025}. 

\section{Projected sensitivity}
\label{app:projected_sensitivity}
In this section, we derive the DM scan rate using a sensor array with the proposed quantum-enhanced scheme, and show how the exclusion of the kinetic mixing $\epsilon$, depends on different system parameters, such as the cavity quality factor, the beamsplitter loss, and the thermal temperature. We rewrite the SNR in Eq.~\eqref{eq:tau_int_approx} as

\begin{align}
    \textrm{SNR} = 2\eta g^2 \tau_{\text{DM}}\sqrt{\frac{\tau_\textrm{tot}}{\gamma_{\uparrow\textrm{eff}}}} N \sqrt{m + 1},
\end{align}
where we accounted for the decoherence effect during the integration and the beamsplitters, and other SPAM errors, using the efficiency parameter $\eta$. Therefore, the total time for reaching a target SNR, $\zeta_\textrm{SNR}$, is given by
\begin{align}
    \tau_\textrm{tot} = \frac{\zeta^2_\textrm{SNR}\omega_\textrm{cav}n_\textrm{th}(\omega_\textrm{cav},T_\textrm{cav})}{4\eta^2g^4\tau_\textrm{DM}^2Q_\textrm{cav}N^2(m+1)},
\end{align}
where $n_\textrm{th}(\omega_\textrm{cav},T_\textrm{cav})=\left(\exp(\hbar\omega_\textrm{cav}/k_\textrm{B}T_\textrm{cav})-1\right)^{-1}$ is the thermal population of the sensor cavity at temperature $T$. 

As demonstrated in Sec.\ref{app:random_walk}, in the regime where the haloscope lifetime far exceeds the dark matter lifetime, a single haloscope tuning step can probe the entire dark matter linewidth $\gamma_\text{DM} = \tau^{-1}_\text{DM}=\omega_\textrm{DM}/Q_\textrm{DM}$. This allows us to write down the scan rate, at frequency $\omega$, as,
\begin{equation}
    \mathcal{R}_\textrm{scan} = \frac{\gamma_\text{DM}}{\tau_\text{tot}}=  16\eta^2\epsilon^4 G^2\rho^2_\mathrm{DM} Q_\textrm{DM}Q_\textrm{cav}\zeta^{-2}_\text{SNR}n^{-1}_\text{th}V^2_\mathrm{cav}(m+1)N^2,
\end{equation}
where we have used $g = \epsilon\,  \sqrt{2 G\rho_\mathrm{DM} V_\mathrm{cav}\, \omega}$ in Eq.~\eqref{eq:DM_cavity_coupling}. Conversely, the kinetic mixing parameter is excluded up to
\begin{align}
     \epsilon(\omega) =\left(\frac{ n_\text{th}\zeta^2_\text{SNR}\omega}{16\eta^2 G^2\rho^2_\mathrm{DM} Q^2_\textrm{DM}Q_\textrm{cav}V^2_\mathrm{cav}\tau_\text{tot}(m+1)N^2}\right)^{1/4},\label{eq:exclusion}
\end{align}

For concreteness, we consider cylindrical Haloscopes with TM$_{010}$ mode being the sensor mode. The mode frequency $\omega_\textrm{cav}$ is determined by the cavity radius $a$~\cite{pozar}, $\omega_\textrm{cav} = 2.4048c/a$, where $c$ is the speed of light. For spectrum control, we set the height $h$ of the haloscope to be twice the diameter, $h=4a$. From this, we may write down the frequency-volume relationship for the cavities,
\begin{align}
    V_\textrm{cav} = \frac{4\pi(2.4048c)^3}{\omega_\textrm{cav}^3}.\label{eq:Vcav}
\end{align}

The TM$_{010}$ mode has an electric field profile of $\hat E(r) = E_0 J_0(2.4048r/a)\hat z$, where $E_0$ is the field amplitude, and $J_0$ is the zero-th order Bessel function of the first kind. Plugging it into Eq.~\eqref{eq:GeometricFactor}, we find the geometric form factor to be

\begin{equation}\label{eq:formfactor}
G= \frac{\left|2\pi h\int_0^a dr\, \hat E(r)\right|^2}{3V_\mathrm{cav}2\pi h\int_0^a dr\, |\hat E(r)|^2 }   = \frac{1}{3}\left(\frac{2}{2.4048}\right)^2 \approx 0.231,
\end{equation}
where the factor of 1/3 accounts for the random polarization of the dark matter field.

\begin{figure*}[t!]
\centering
\includegraphics[width=0.5
\textwidth]{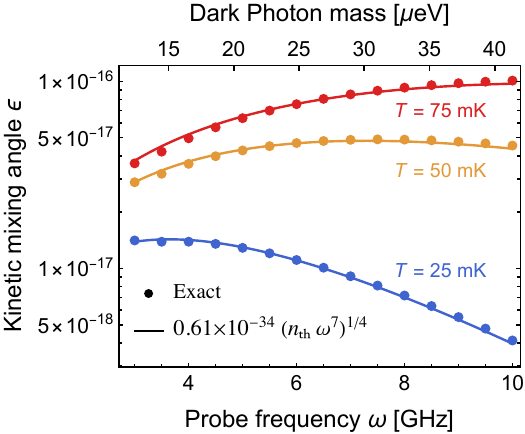}
\caption{\justifying
\textbf{Kinetic mixing angle exclusion for different Dark Photon mass at different temperatures.} The numerical results (circles) can be well fitted by the analytical function of $0.61\times10^{-34}\left(n_\textrm{th}\,\omega^7\right)^{1/4}$ (solid lines). Lower temperature corresponds to a lower heating rate, which results in an overall higher sensitivity.}
\label{fig:exclusion_vs_temp}
\end{figure*}

In Fig.~\ref{fig:exclusion_vs_temp}, we plot the exclusion of the kinetic mixing angle at $90\%$ confidence level (corresponding to $\zeta_\textrm{SNR} = 1.28$), for a four-cavity array with five-photon initial Fock state ($N=4,\, m=5$), in a total integration time of $\tau_\textrm{tot}=\SI{10}{\second}$. We set $\rho_{\mathrm{DM}} = \SI{0.45}{\giga\electronvolt\per\cubic\centi\metre}$, $Q_\textrm{DM}=10^6$, $Q_\textrm{cav}=10^8$. We assume moderate beamsplitter fidelity $F_\textrm{BS}=99\%$, and consider a SPAM time of \SI{20}{\micro\second} that allows for repeated measurement in each cycle. The calculation is done via the numerical approach outlined in Sec.~\ref{app:simulation}. We consider three different cavity temperatures, $T_\textrm{cav} = \SI{25}{\milli\kelvin}$, \SI{50}{\milli\kelvin} and \SI{75}{\milli\kelvin}. Combining Eq.~\eqref{eq:exclusion} and Eq.~\eqref{eq:Vcav}, we have $\epsilon(\omega)\propto (n_\text{th}\omega^7)^{1/4}$. The concaveness of $\epsilon(\omega)$ can be well explained by the competition between $n_\text{th}$ (the thermal population) and $\omega^7$ (primarily due to the volume of the cavity), which can be see as the good agreement between the numerical result and the fit using $(n_\text{th}\omega^7)^{1/4}$.

\end{document}